\documentclass[useAMS,usenatbib]{mn2e}  
\usepackage{graphicx,amssymb,afterpage}
\begin{document}
   \title{Rotation measure synthesis revisited
}

   \author[Schnitzeler \& Lee]
          {D.H.F.M. Schnitzeler$^{1}$\thanks{Schnitzeler@mpifr-bonn.mpg.de}, 
           K.J. Lee$^{2,1}$\\
          $^1$ Max Planck Institut f\"ur Radioastronomie, 53121 Bonn, Germany\\
          $^2$ Kavli institute for radio astronomy and astrophysics, Peking University, Beijing 100871, P.R.China\\
          }

   \date{}

   \pagerange{}\pubyear{}

   \maketitle
 
\begin{abstract}
We re-formulate rotation measure (RM) synthesis for data sets with discrete frequency channels and an arbitrary channel response function.
The most commonly used version of the formalism by Brentjens \& De Bruyn assumes a top-hat response function in wavelength squared, while real data sets can often be approximated better with a top-hat in frequency.
We simulate mock data sets for various source geometries, using a top-hat response function in frequency, and we compare the quality of the RM spectra that are found with both formalisms. 
We include the response function of the simulated data to calculate exact RM spectra using our formalism.
We show that the formalism by Brentjens \& De Bruyn produces accurate results even if depolarization at the lowest frequencies in the observing band is severe.
If RMs are large, our formalism reconstructs the emitted signal more accurately, with a higher amplitude and (in most cases) a narrower RM spread function.
Our formalism can also detect sources with larger (absolute) RMs for a given sensitivity level of the observations. 
\end{abstract}
\begin{keywords} Polarization -- Magnetic fields -- Techniques: polarimetric -- Techniques: spectroscopic -- Methods: analytical
\end{keywords}

%

\section{Introduction}\label{sec-introduction}
Rotation measure synthesis, as developed by \citet{burn1966} and \citet[][`B05']{brentjens2005}, identifies the observed monochromatic polarization vector as a Fourier transform between rotation measure (RM) space and wavelength squared ($\lambda^2$) space, and uses the inverse transform to reconstruct the emitted signal:
\begin{eqnarray}
\bmath{P}\left(\lambda^2\right) & = & \int_{-\infty}^\infty \bmath{P}\left(\mathrm{RM}\right)_\mathrm{true}\mathrm{e}^{2\mathrm{iRM}\lambda^2}\mathrm{dRM}\, ,\ \mathrm{and} \label{rot_classic} \\
\bmath{P}\left(\mathrm{RM}\right) & = & \frac{\int_{-\infty}^{\infty} l\left(\lambda^2\right) \bmath{P}\left(\lambda^2\right)\mathrm{e}^{-2\mathrm{iRM}\lambda^2}\mathrm{d}\lambda^2}{\int_{-\infty}^{\infty} l\left(\lambda^2\right)\mathrm{d}\lambda^2}\, , 
\label{derot_classic}
\end{eqnarray}
\noindent
where $\bmath{P}= Q+\mathrm{i}U$ is the complex polarization vector that has Stokes $Q$ and $U$ as its real and imaginary parts. $\bmath{P}\left(\mathrm{RM}\right)_\mathrm{true}$ and $\bmath{P}\left(\mathrm{RM}\right)$ indicate the true (emitted) and reconstructed RM spectra. $l\left(\lambda^2\right)$ is the response function that expresses the relative weight of measurements at $\lambda^2$.  
Frequency channels have a finite width, and the RM spectrum is normally calculated using
\begin{eqnarray}
\bmath{P}\left(\mathrm{RM'}\right) & = & \frac{1}{N_\mathrm{c}}\sum_{j=1}^{N_\mathrm{c}} \bmath{P}\left(\lambda_{\mathrm{c},j}^2\right)\mathrm{e}^{-2\mathrm{i}\mathrm{RM'}\lambda^2_{\mathrm{c},j}}\, ,
\label{derot_sum}
\end{eqnarray}
for a trial rotation measure $\mathrm{RM}'$. There are $N_\mathrm{c}$ channels, `$j$' is the channel index, and $\lambda^2_{\mathrm{c},j}$
is the central wavelength squared of channel $j$, which lies between $\lambda^2_j$ and  $\lambda^2_{j+1}$.
Equation~\ref{derot_sum} is exact if each channel has a top-hat response in wavelength squared.
However, the response functions of real data sets can often be approximated more accurately by a top-hat in frequency. 
In this Letter we investigate how important this difference between the two formalisms is.

In Section~\ref{sec-formalism} we formulate RM synthesis for frequency channels with an arbitrary response function.
In Section~\ref{sec-recon} we qualitatively and quantitatively analyse how Equation~\ref{derot_sum} and our new formalism reconstruct RM spectra for a range of simulated source types, observing frequencies, and frequency channel widths.

\section{Formalism}\label{sec-formalism}
To understand how RM spectra can be calculated for frequency channels with an arbitrary response function we first formulate the RM synthesis formalism for channels with a top-hat response in wavelength squared, given by Equation~\ref{derot_sum}. Then we generalise this formalism. 
The integral of the response function of each channel is equal to one, which ensures that the flux density which is measured in a channel is equal to the mean flux density of the source across the channel. 
Since $l\left(\lambda^2\right) = \sum_{j=1}^{N_\mathrm{c}} l_j\left(\lambda^2\right)$, it follows that $\int_{-\infty}^{\infty} l\left(\lambda^2\right) \mathrm{d}\lambda^2 = N_\mathrm{c}$.
The response functions can extend beyond the channel edges, and overlap.
To remove spectral index effects we divide Stokes $Q$ and $U$ and the amplitude of the polarization vector $\left|\bmath{P}\right| \equiv P$ by Stokes $I$, and indicate the resulting quantities with lower-case letters: $q$, $u$, and ${p}$.

Consider a source that emits at a single RM.
The observed net polarization vector of a single channel is
\begin{eqnarray}
\bmath{p}\left(\lambda^2_{\mathrm{c},j}\right) & \equiv & \int_{-\infty}^{\infty} l_j\left(\lambda^2\right)\bmath{p}\left(\lambda^2\right)\mathrm{d}\lambda^2\, ,
\label{channel_depol_single_RM}
\end{eqnarray}
where $\lambda^2_{\mathrm{c},j}= \int_{-\infty}^{\infty} l_j\left(\lambda^2\right)\lambda^2\mathrm{d}\lambda^2$ is the weighted mean $\lambda^2$ of the channel, and $\bmath{p}\left(\lambda^2\right)$ follows from Equation~\ref{rot_classic}.
RM spectra can be calculated by aligning the net polarization vectors of the different channels for an assumed $\mathrm{RM}'$, and adding these vectors:
\begin{eqnarray}
\lefteqn{\bmath{p}\left(\mathrm{RM}'\right)  \equiv
\frac{1}{N_\mathrm{c}} \sum_\mathrm{j=1}^{N_\mathrm{c}}
\bmath{p}\left(\lambda^2_\mathrm{c,j}\right) \hat{\bmath{v}}_j\left(\mathrm{RM}'\right)_\mathrm{derot}\, . }
\label{RM_derot_discrete_lambdatwo}
\end{eqnarray}
We have introduced the net derotation vector $\bmath{v}_j\left(\mathrm{RM}'\right)_\mathrm{derot} $, which corrects for the net amount of Faraday rotation of each channel:
\begin{eqnarray} 
\bmath{v}_j\left(\mathrm{RM}'\right)_\mathrm{derot} 
& \equiv &
\int_{-\infty}^{\infty} l_j\left(\lambda^2\right) \mathrm{e}^{-2\mathrm{iRM}'\lambda^2}\mathrm{d}\lambda^2\, . 
\label{net_derot_lambdatwo}
\end{eqnarray}
Equation~\ref{RM_derot_discrete_lambdatwo} requires that the net derotation vector is normalised, 
$
\hat{\bmath{v}}_j\left(\mathrm{RM}'\right)_\mathrm{derot} \equiv \bmath{v}_j\left(\mathrm{RM}'\right)_\mathrm{derot}/\left| \bmath{v}_j\left(\mathrm{RM}'\right)_\mathrm{derot}\right| .
$

For frequency channels with a top-hat response function in wavelength squared, which is uniform within the channel and zero outside it, the solution to Equation~\ref{channel_depol_single_RM} is
\begin{eqnarray}
\bmath{p}\left(\lambda^2_{\mathrm{c},j}\right) =  \bmath{p}\left(\mathrm{RM}\right)_\mathrm{true}\mathrm{sinc}\left(\mathrm{RM}\delta\lambda^2\right)\mathrm{e}^{2\mathrm{iRM\lambda^2_{\mathrm{c},j}}}\, .
\label{channel_depol_single_RM_tophat}
\end{eqnarray}
In this case the channel width $\delta\lambda^2$ = $\lambda^2_{j+1} - \lambda^2_j$, 
$\lambda^2_{\mathrm{c},j} = \left(\lambda_j^2+\lambda_{j+1}^2\right)/2$, and $\mathrm{sinc}\left(x\right) \equiv \mathrm{sin}\left(x\right)/x$.
The normalised net derotation vector for this channel response function is equal to
\begin{eqnarray}
\lefteqn{
\hat{\bmath{v}}_j\left(\mathrm{RM}'\right)_\mathrm{derot} = \mathrm{e}^{-2\mathrm{iRM}'\lambda^2_{\mathrm{c},j}}\, . 
}
\nonumber
\end{eqnarray}
Therefore, Equation~\ref{RM_derot_discrete_lambdatwo} simplifies to Equation~\ref{derot_sum} if frequency channels have a top-hat response in wavelength squared.
The amplitude of the reconstructed signal will be smaller than the amplitude of the emitted signal $\bmath{p}\left(\mathrm{RM}\right)_\mathrm{true}$ because of the sinc factor in Equation~\ref{channel_depol_single_RM_tophat}.

Now we formulate expressions for calculating RM spectra for channels with an arbitrary response in frequency. The observed net polarization vector for a source that emits at a single RM is given by
\begin{eqnarray}
\bmath{p}\left(\nu_{\mathrm{c},j}\right) = 
\int_{-\infty}^{\infty} w_j\left(\nu\right) \bmath{p}\left(\mathrm{RM}\right)_\mathrm{true}
\mathrm{e}^{2\mathrm{iRM}\left(\mathrm{c}/\nu\right)^2}\mathrm{d}\nu\, , \, \label{weighted_channel_depol_single_RM_frequency}
\end{eqnarray}
where $\nu$ indicates the observing frequency, `$\mathrm{c}$' the speed of light, and $w_j\left(\nu\right)$ the (normalised) response function in frequency of channel $j$. $\nu_{\mathrm{c},j}$ is the weighted mean frequency of channel $j$, $\nu_{\mathrm{c},j} = \int_{-\infty}^{\infty} w_j\left(\nu\right)\nu\mathrm{d}\nu$.
RM spectra can be calculated by aligning the measured polarization vectors for an assumed trial $\mathrm{RM}'$ and then summing the derotated vectors, similar to Equation~\ref{RM_derot_discrete_lambdatwo}:
\begin{eqnarray}
\lefteqn{\bmath{p}\left(\mathrm{RM}'\right)  = 
\frac{1}{N_\mathrm{c}} \sum_\mathrm{j=1}^{N_\mathrm{c}}
\bmath{p}\left(\nu_{\mathrm{c},j}\right) \hat{\bmath{v}}_j\left(\mathrm{RM}'\right)_\mathrm{derot}\, , }
\label{RM_derot_discrete}
\end{eqnarray}
where the net derotation vector for channel $j$ (which has not been normalised) is defined as
\begin{eqnarray}
\bmath{v}_j\left(\mathrm{RM}'\right)_\mathrm{derot} \equiv  
\int_{-\infty}^{\infty} w_j\left(\nu\right) \mathrm{e}^{-2\mathrm{iRM}'\left(\mathrm{c}/\nu\right)^2}\mathrm{d}\nu\, \label{weighted_net_derot_frequency}\, .
\end{eqnarray}

In the remainder of this Letter we will consider frequency channels with top-hat weighting functions in frequency. In that case Equations~\ref{weighted_channel_depol_single_RM_frequency} and \ref{weighted_net_derot_frequency} simplify to
\begin{eqnarray}
\bmath{p}\left(\nu_{\mathrm{c},j}\right) = 
\frac{1}{\delta\nu} 
\int_{\nu_j}^{\nu_{j+1}} \bmath{p}\left(\mathrm{RM}\right)_\mathrm{true}
\mathrm{e}^{2\mathrm{iRM}\left(\mathrm{c}/\nu\right)^2}\mathrm{d}\nu\, , \mathrm{and}
\label{channel_depol_single_RM_frequency}
\end{eqnarray}
\begin{eqnarray}
\bmath{v}_j\left(\mathrm{RM}'\right)_\mathrm{derot}  = 
\frac{1}{\delta\nu} 
\int_{\nu_j}^{\nu_{j+1}} \mathrm{e}^{-2\mathrm{iRM}'\left(\mathrm{c}/\nu\right)^2}\mathrm{d}\nu\, ,
\label{net_derot_frequency}
\end{eqnarray}
where $\delta\nu = \nu_{j+1}-\nu_j$.
The integral in Equation~\ref{net_derot_frequency} can be evaluated in the complex plane, where we use the solution $\sqrt{\mathrm{i}} = \left(1+\mathrm{i}\right)/\sqrt{2}$. In its indefinite form: 
\begin{eqnarray}
\lefteqn{
\int
\mathrm{e}^{-2\mathrm{iRM}'\left(\mathrm{c}/\nu\right)^2}\mathrm{d}\nu =
\mathrm{e}^{-2\mathrm{iRM}'\left(c/\nu\right)^2}\nu 
 + \mathrm{c}\sqrt{\left|\mathrm{RM}'\right|\pi} } \nonumber \\
&& \left(\mathrm{i} + \mathrm{sign}\left(\mathrm{RM}'\right) \right)
\mathrm{erf}\left[\sqrt{\left|\mathrm{RM}'\right|}\frac{\mathrm{c}}{\nu}\left(\mathrm{i}+\mathrm{sign}\left(\mathrm{RM}'\right) \right) \right]
\label{eminus}
\end{eqnarray}
\noindent
(omitting the integration constant). The error function `erf' is defined as
\begin{eqnarray}
\mathrm{erf}\left(x\right) \equiv \frac{2}{\sqrt{\pi}}\int_0^x \mathrm{e}^{-t^2}\mathrm{d}t\, .
\nonumber
\end{eqnarray}
\noindent
The error function can be written in terms of the lower incomplete gamma function, $\mathrm{erf}\left(x\right) \propto \gamma\left(1/2,x^2\right)$,
which makes it possible to evaluate the error function when its argument is complex-valued. `sign(x)' returns the sign of the argument, and 0 when the argument is zero.

\begin{figure*}
    \resizebox{0.32\hsize}{!}{\includegraphics[width=\linewidth]{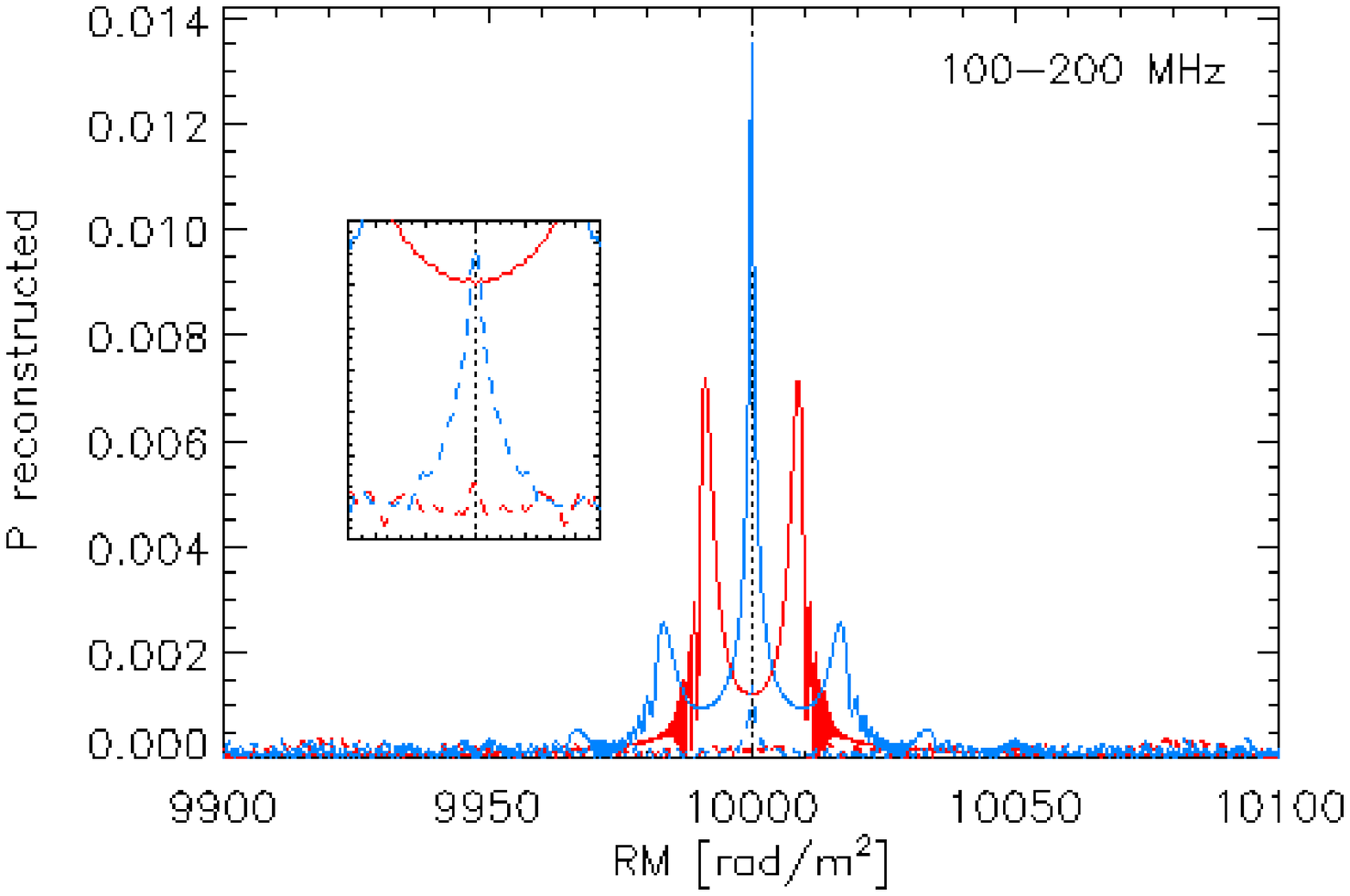}} 
    \resizebox{0.31\hsize}{!}{\includegraphics[width=\linewidth]{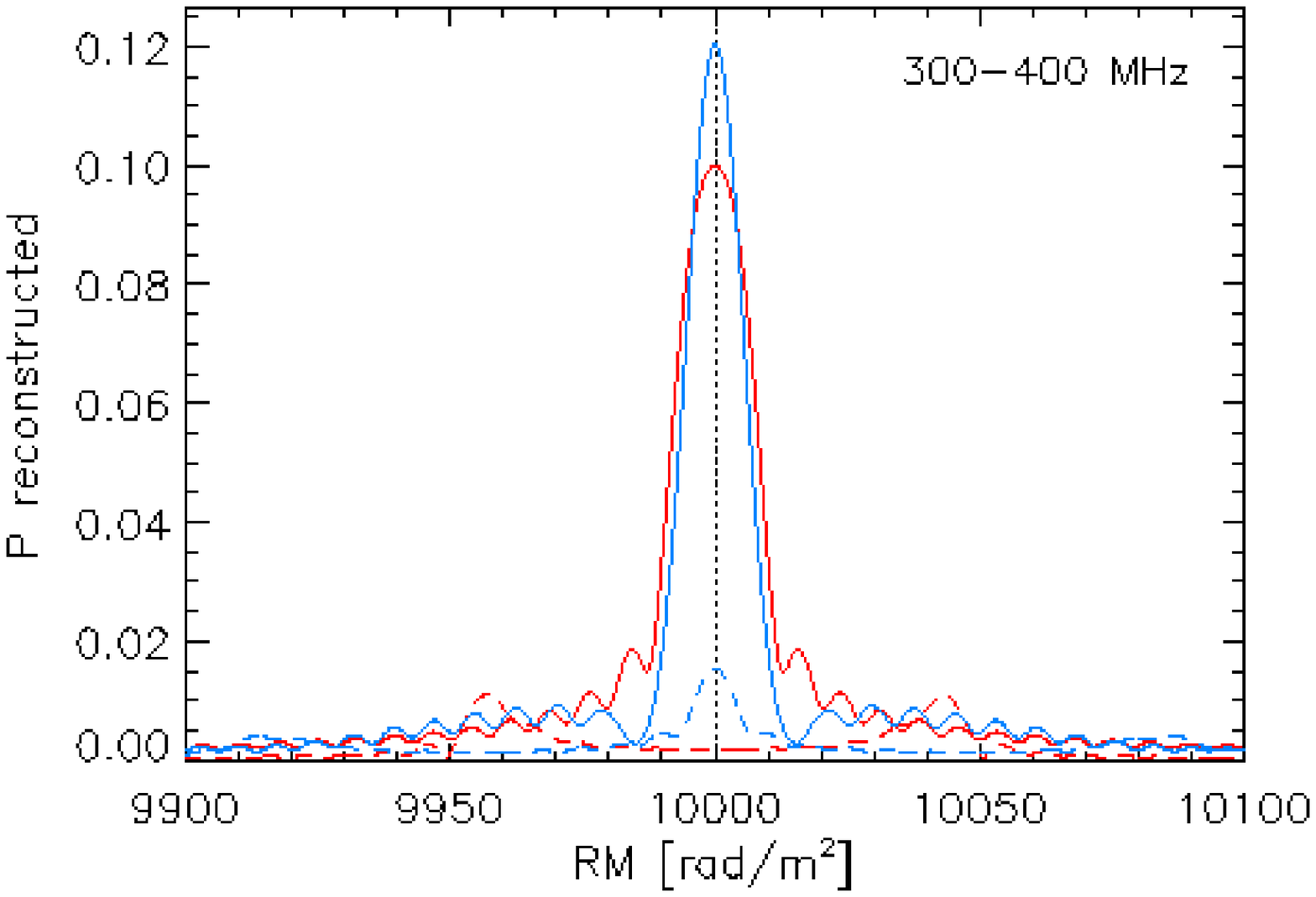}} 
    \resizebox{0.30\hsize}{!}{\includegraphics[width=\linewidth]{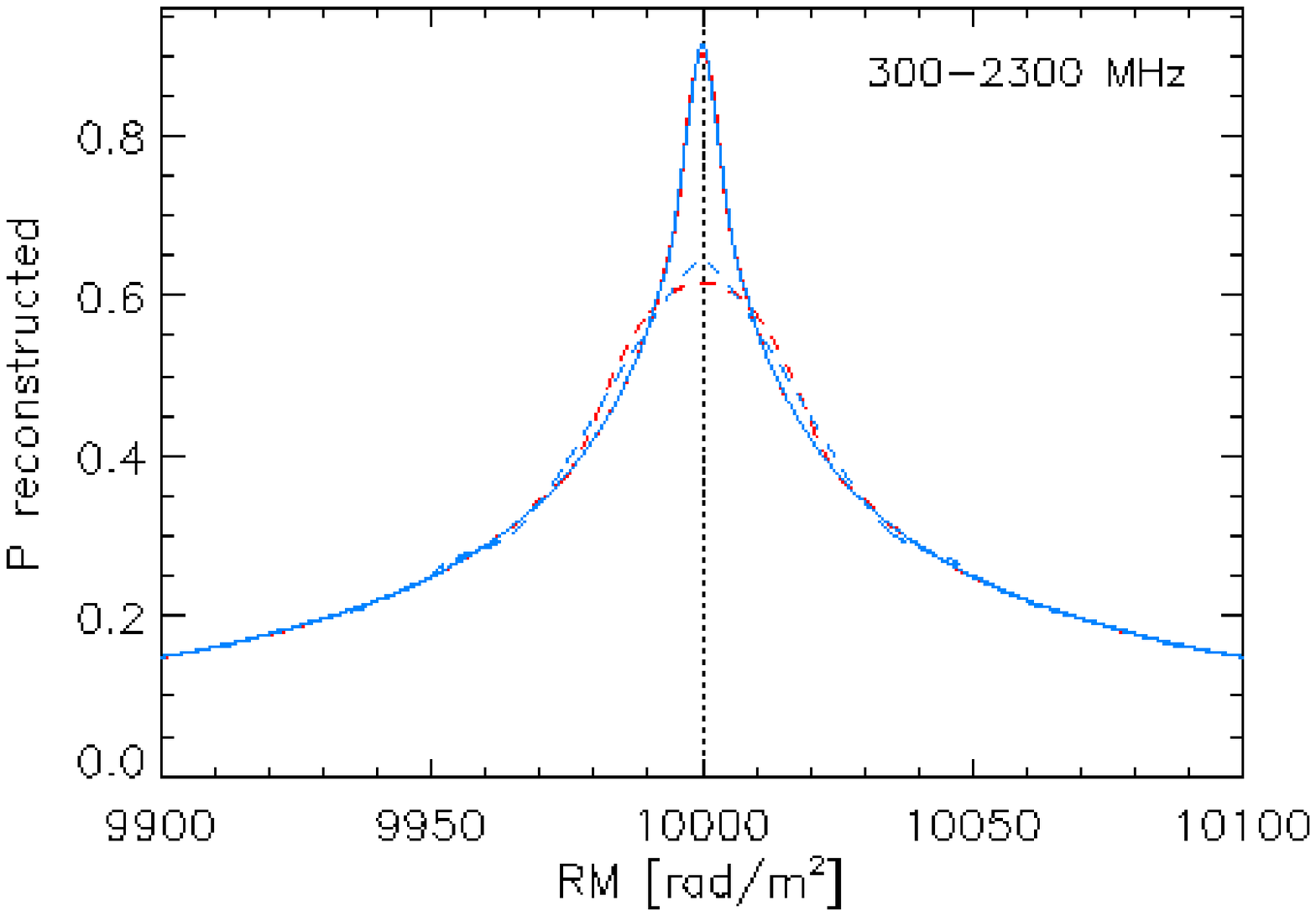}} 
%
%
    \resizebox{0.32\hsize}{!}{\includegraphics[width=\linewidth]{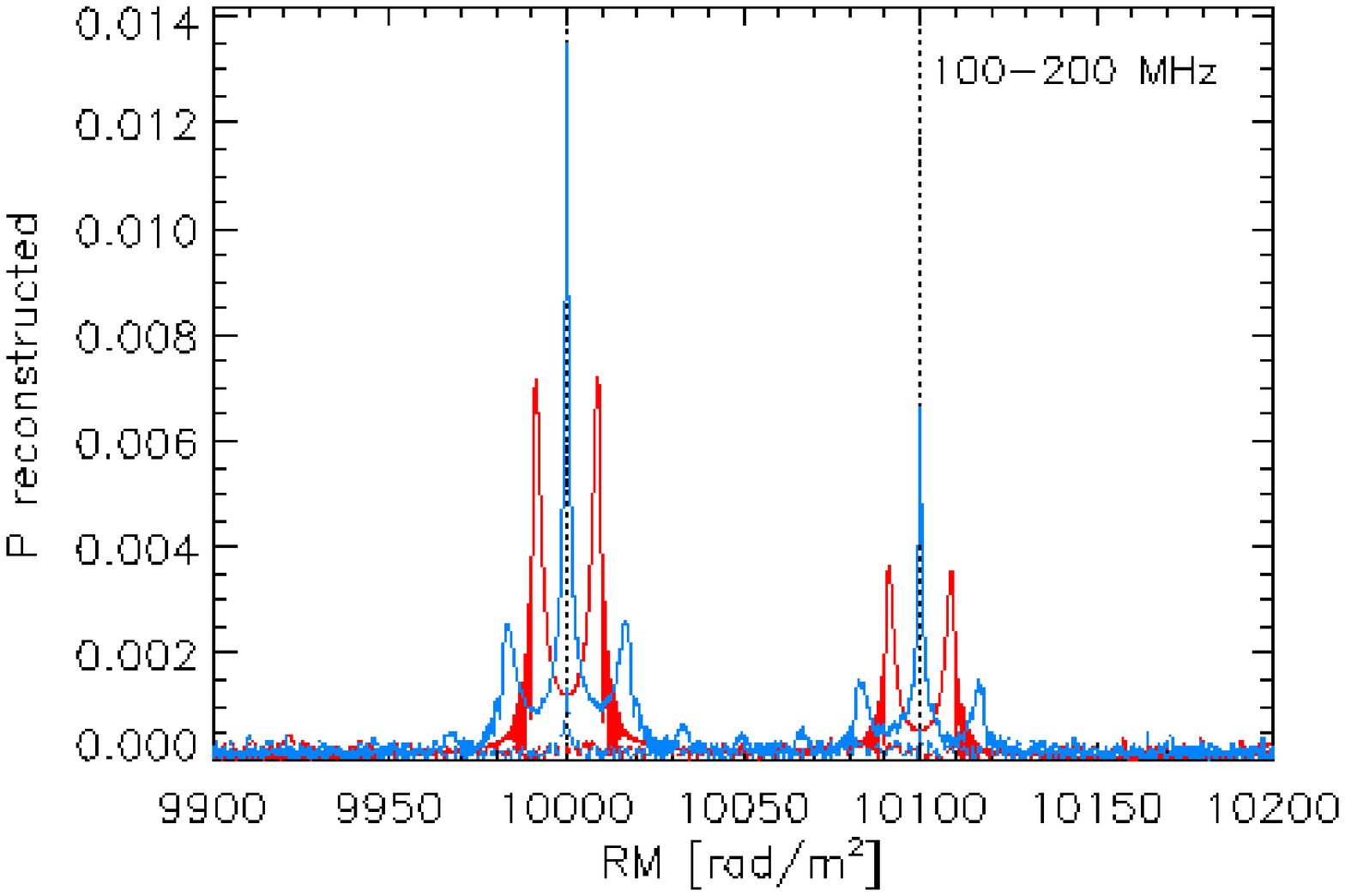}} 
    \resizebox{0.31\hsize}{!}{\includegraphics[width=\linewidth]{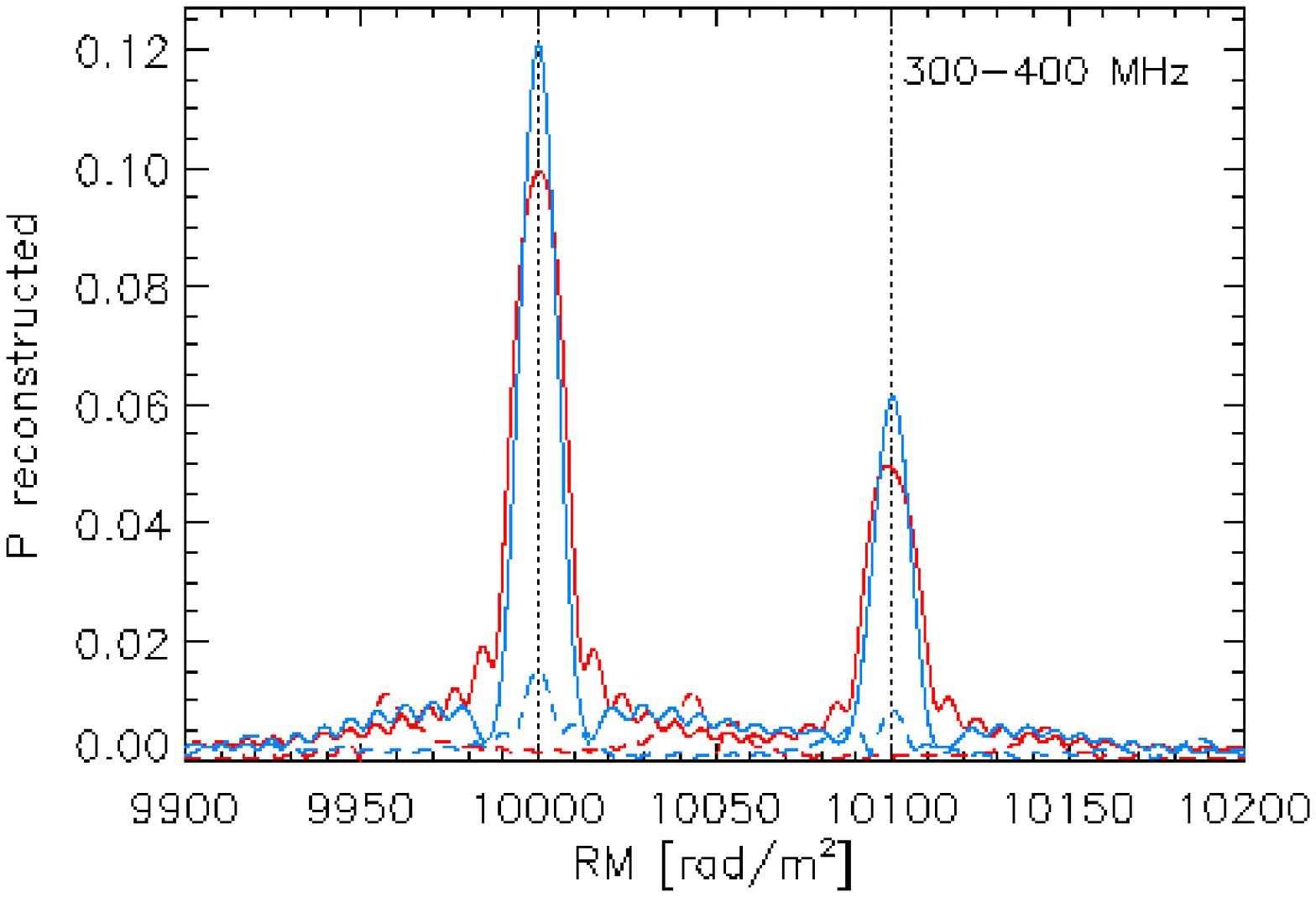}} 
    \resizebox{0.30\hsize}{!}{\includegraphics[width=\linewidth]{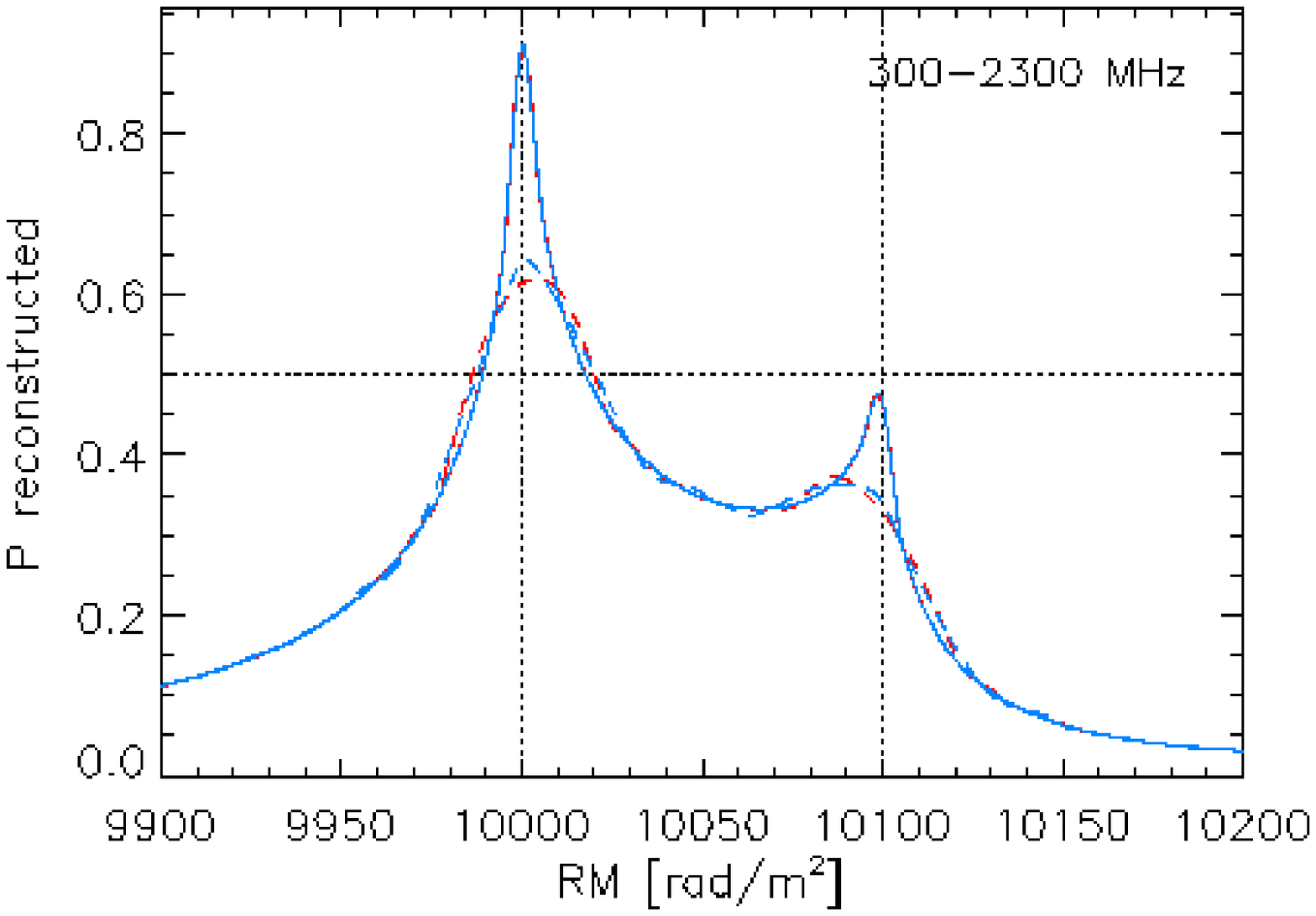}} 
%
%
    \resizebox{0.315\hsize}{!}{\includegraphics[width=\linewidth]{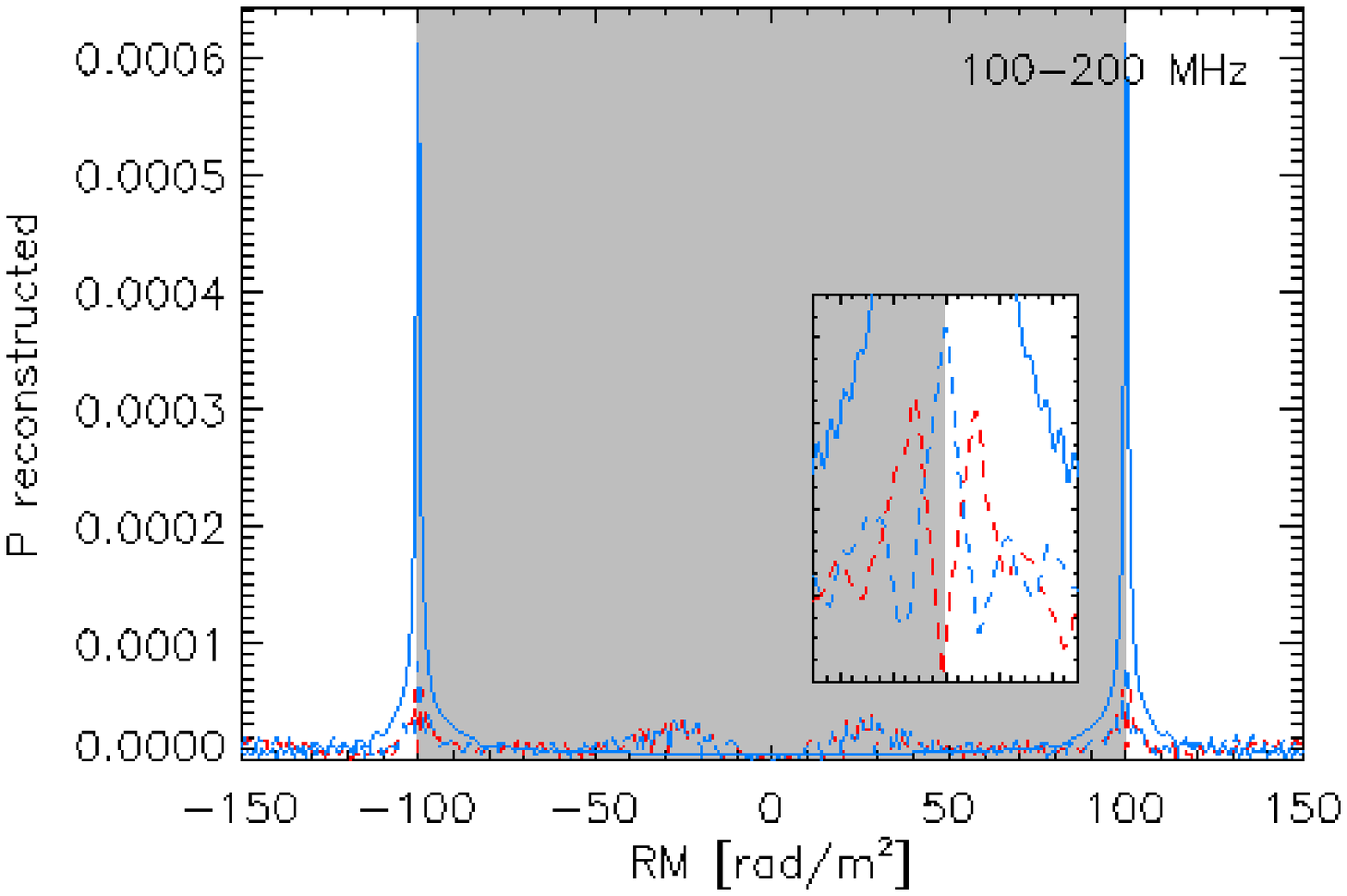}} 
    \resizebox{0.31\hsize}{!}{\includegraphics[width=\linewidth]{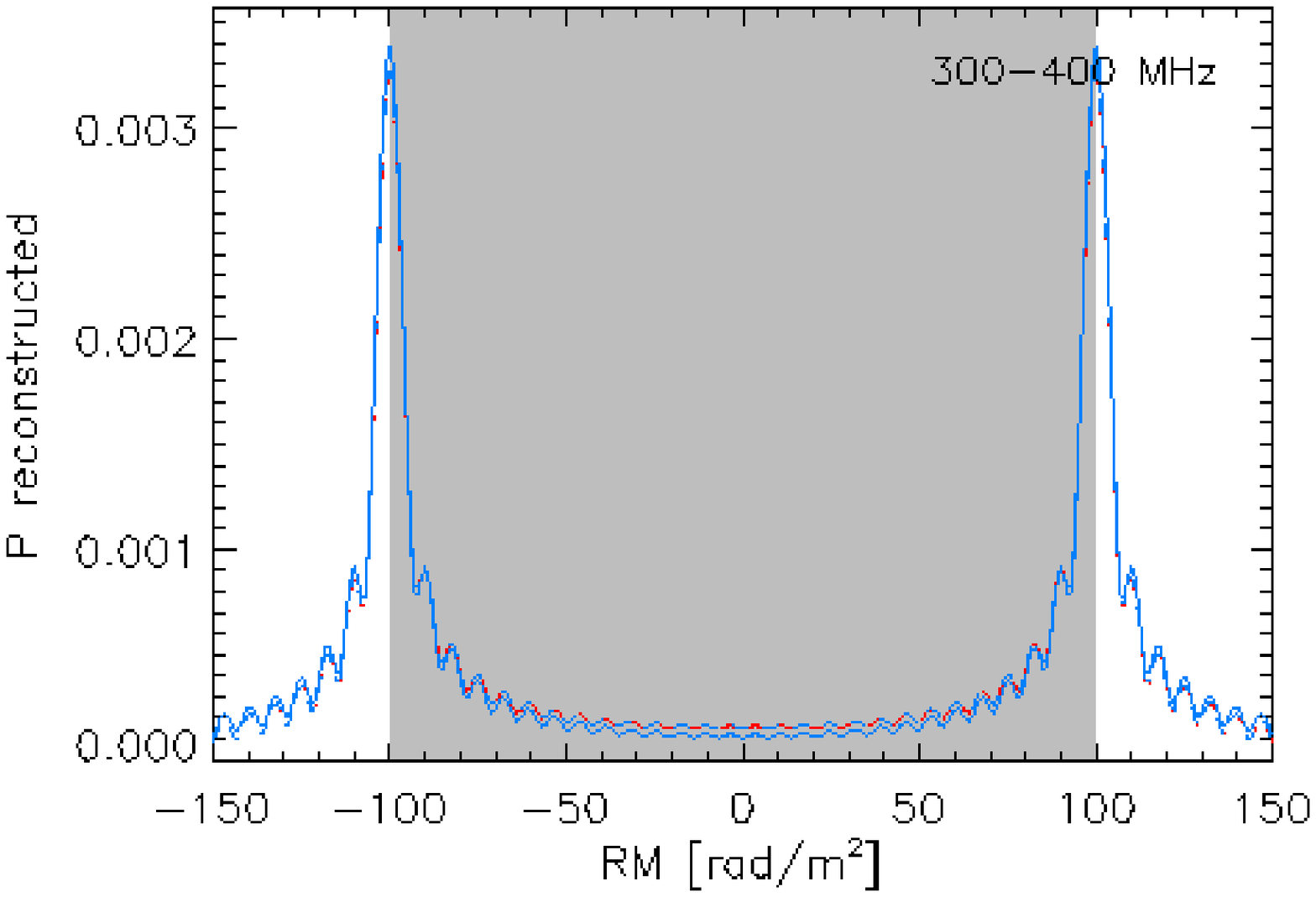}} 
    \resizebox{0.30\hsize}{!}{\includegraphics[width=\linewidth]{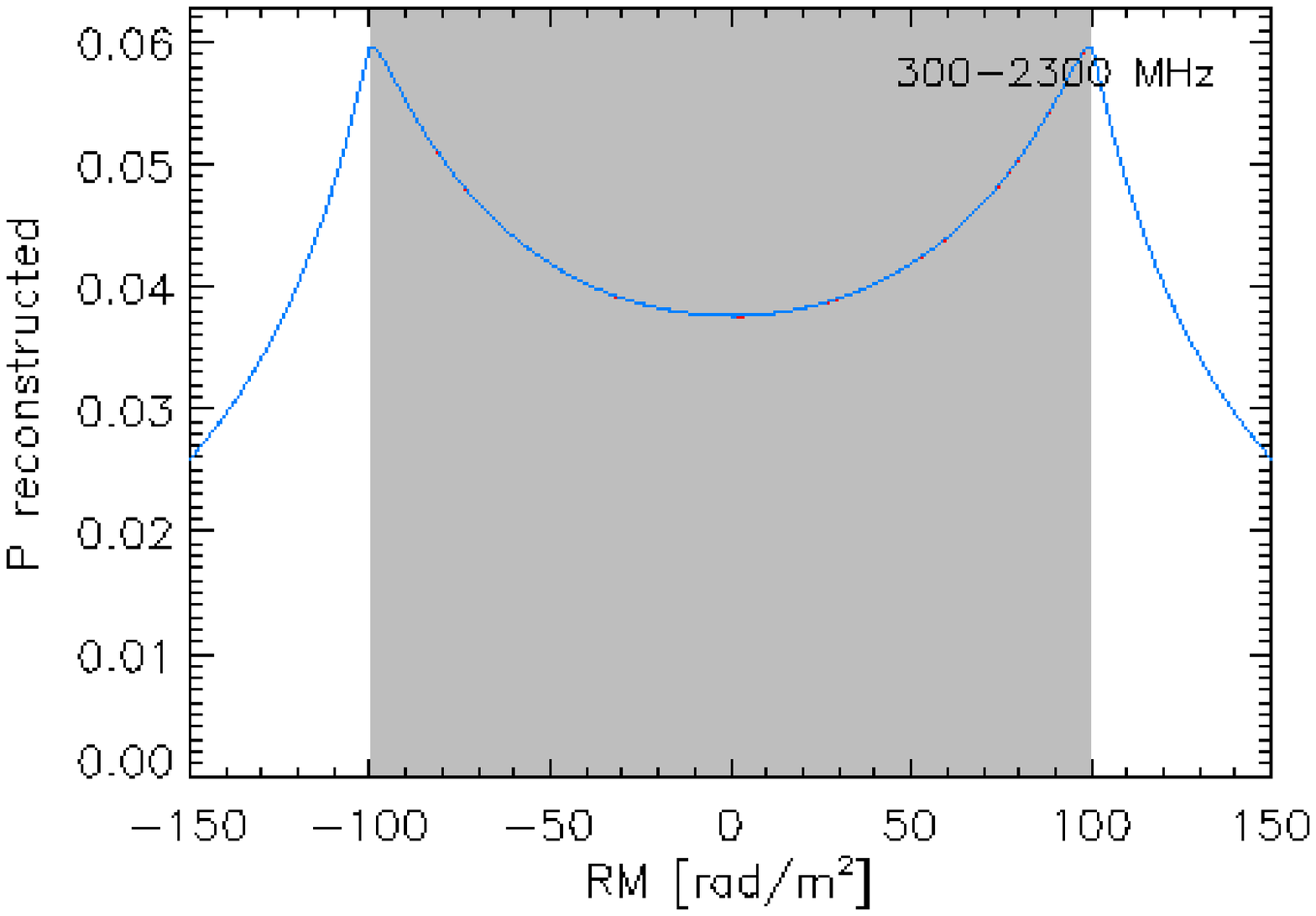}} 
%
%
    \resizebox{0.30\hsize}{!}{\includegraphics[width=\linewidth]{empty.ps}} 
    \resizebox{0.31\hsize}{!}{\includegraphics[width=\linewidth]{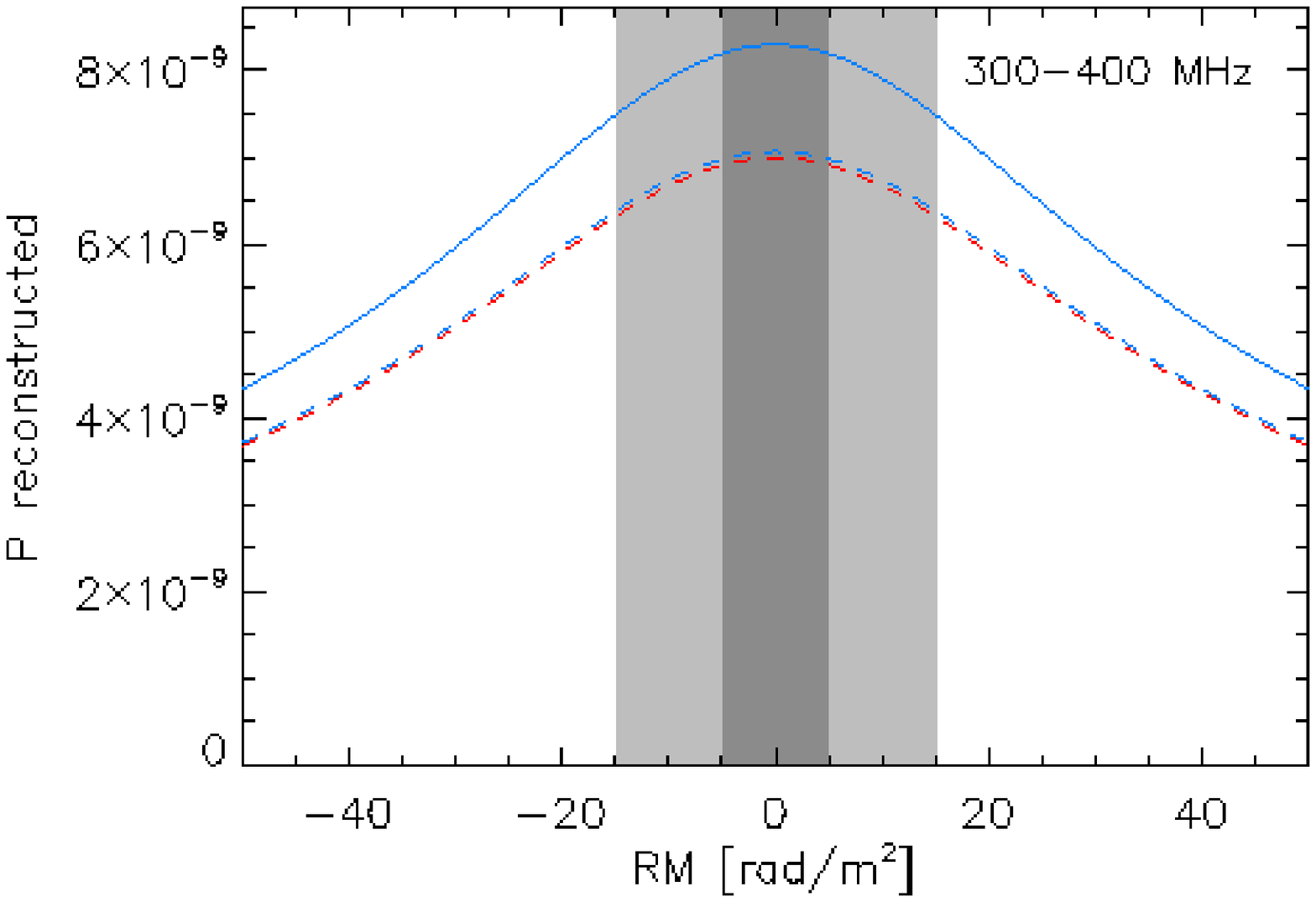}} 
    \resizebox{0.29\hsize}{!}{\includegraphics[width=\linewidth]{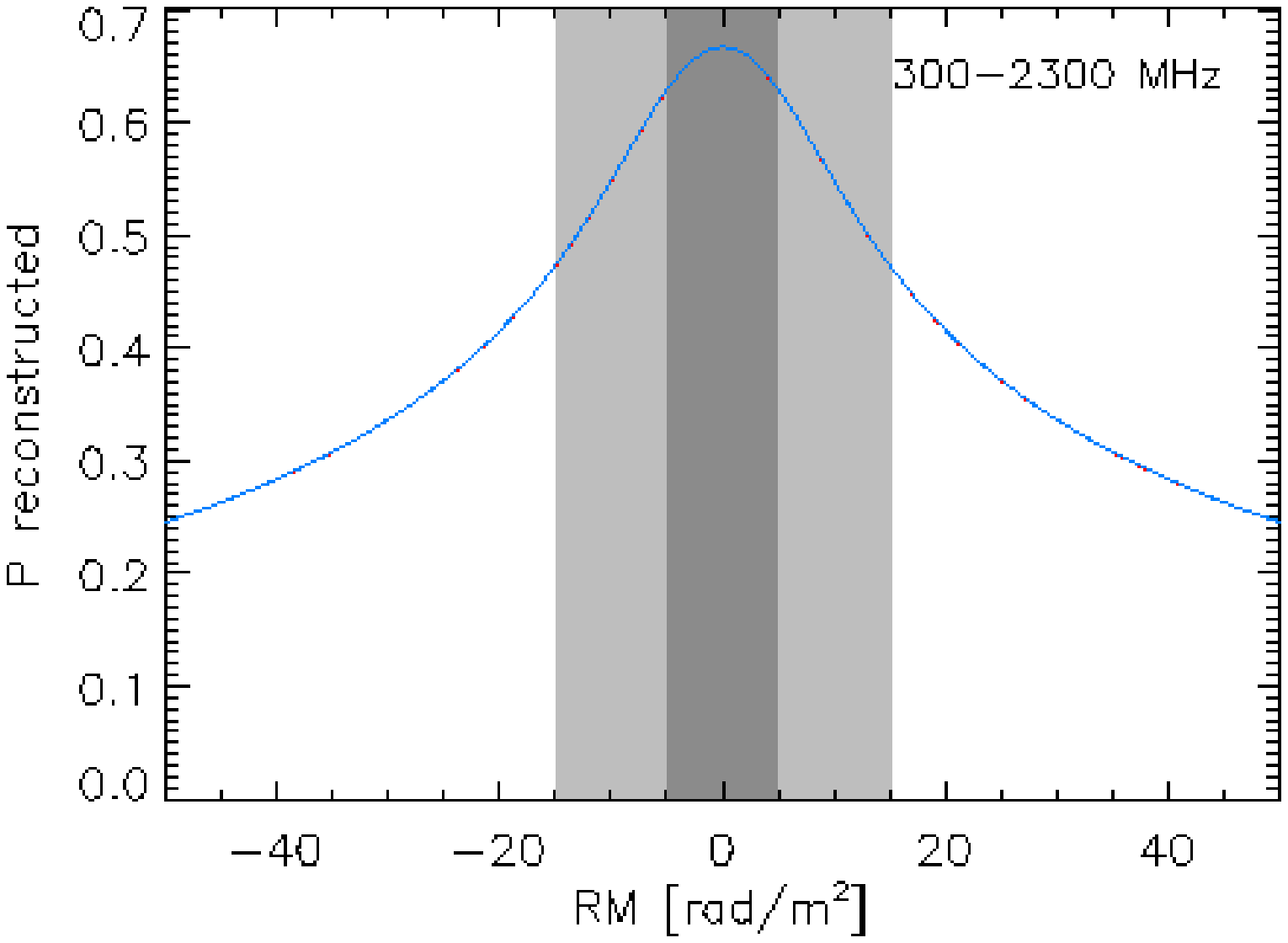}} 
\caption{Reconstructed RM spectra using either our reconstruction method with the normalised version of Equation~\ref{net_derot_frequency} (blue curves) or Equation~\ref{derot_sum} (red curves). From top to bottom: source emits at a single RM (\emph{first row of panels}), two RMs (\emph{second row}), a source with a rectangular $\left|\bmath{p}\left(\mathrm{RM}\right)\right|$ distribution (\emph{third row}), and a source with a Gaussian distribution ($\sigma_\mathrm{RM}$ = 5~rad~m$^{-2}$; \emph{bottom row}). All panels have the same layout, and show the polarized flux density of the reconstructed RM spectrum as a function of RM in rad~m$^{-2}$. In all cases the maximum of the emitted spectrum is set to 1. The frequency range that was used in the reconstruction is shown in the upper-right of each panel. Solid lines: 0.1 MHz channels. Dashed lines: 1 MHz channels. Insets highlight the behaviour of the RM spectra when the data are sampled with 1 MHz channels. The vertical dotted lines indicate the RM of the components (first two rows of panels), while the grey background in the third row shows the range in RM with polarized emission. The light/dark grey regions in the fourth row show the 3-sigma and 1-sigma intervals, respectively, of the Gaussian emission region. 
The peak flux density in the RM spectrum of the Gaussian source is $< 10^{-45}$ for observations in the 100 to 200 MHz band, and we do not show it.
}
\label{RM-spec.fig}
\end{figure*}

\section{Reconstructed RM spectra}\label{sec-recon}
In this section we compare the RM spectra that are reconstructed using Equation~\ref{derot_sum} and the new formalism we propose using the same mock observations as input.
In Fig.~\ref{RM-spec.fig} we compare RM spectra that are found using the two formalisms for four source geometries: a source that emits at a single RM, a source that emits at two RMs, a source that emits a constant signal over a continuous range in RM, which is also known as a Burn slab \citep{burn1966}, and, finally, a source with a Gaussian emission profile that illuminates a linear RM gradient.
We simulate observations of Stokes $q$ and $u$ with discrete channels and a top-hat response function in frequency using Equation~\ref{channel_depol_single_RM_frequency}.
These simulated observations are then derotated and summed using Equation~\ref{derot_sum}. For each frequency channel we convert the frequencies at the channel edges to $\lambda^2$, and calculate $\lambda_\mathrm{c}^2$ and $\delta\lambda^2$ from these, sidestepping the approximations used in equations~34 and 35 in B05\footnote{Equation~34 in B05 uses a second-order Taylor expansion to express $\lambda_\mathrm{c}^2$ in terms of the width of the frequency channel and its mean frequency, while equation~35 in B05 uses a third-order Taylor expansion to express $\delta\lambda^2$ in terms of these quantities.}. 
To calculate RM spectra with our formalism we use the normalised version of Equation~\ref{net_derot_frequency}, written out as Equation~\ref{eminus}.

\begin{figure*}
    \resizebox{0.32\hsize}{!}{\includegraphics[width=\linewidth]{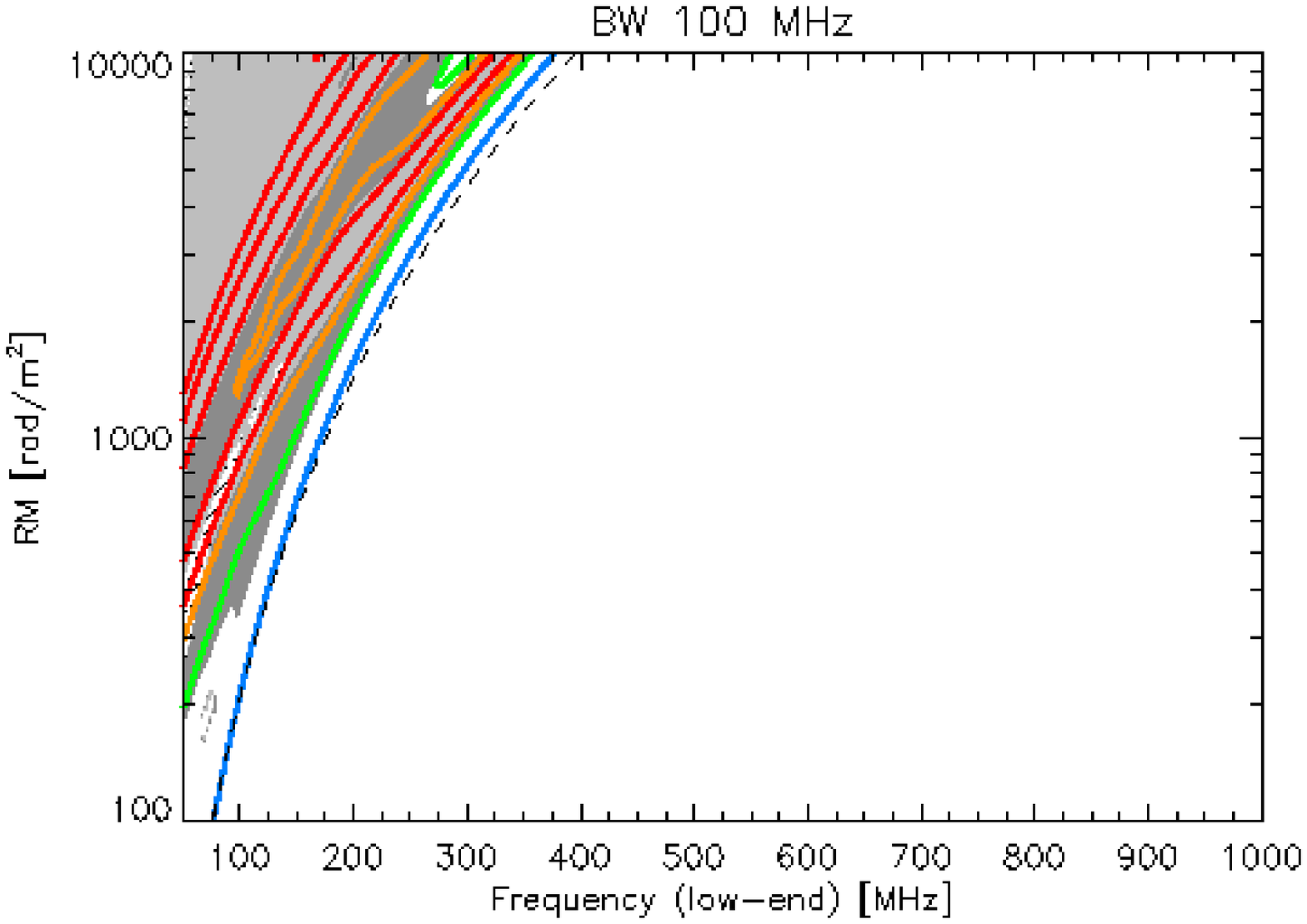}} 
    \vspace{-1ex}
    \resizebox{0.32\hsize}{!}{\includegraphics[width=\linewidth]{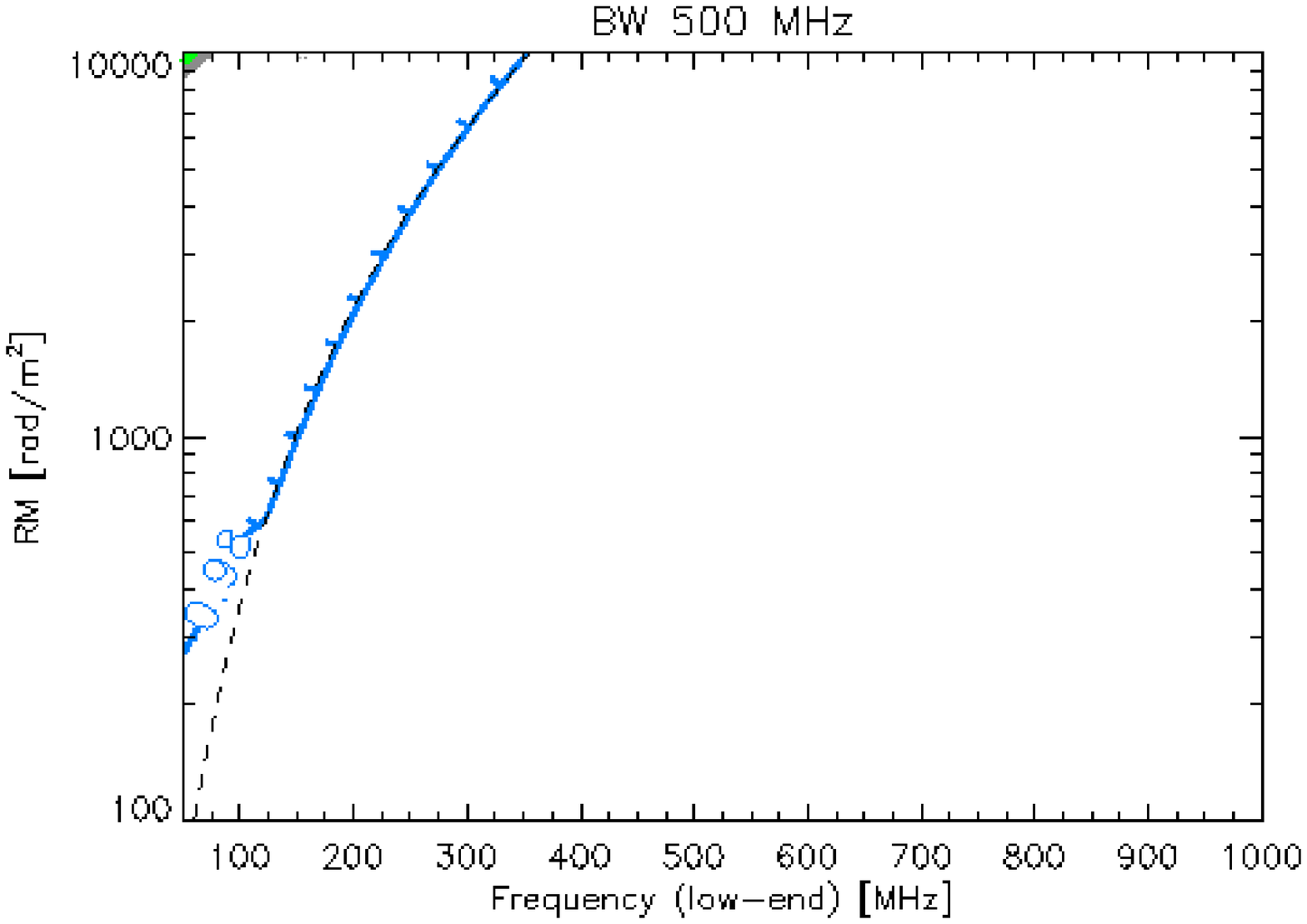}} 
    \vspace{-1ex}
    \resizebox{0.32\hsize}{!}{\includegraphics[width=\linewidth]{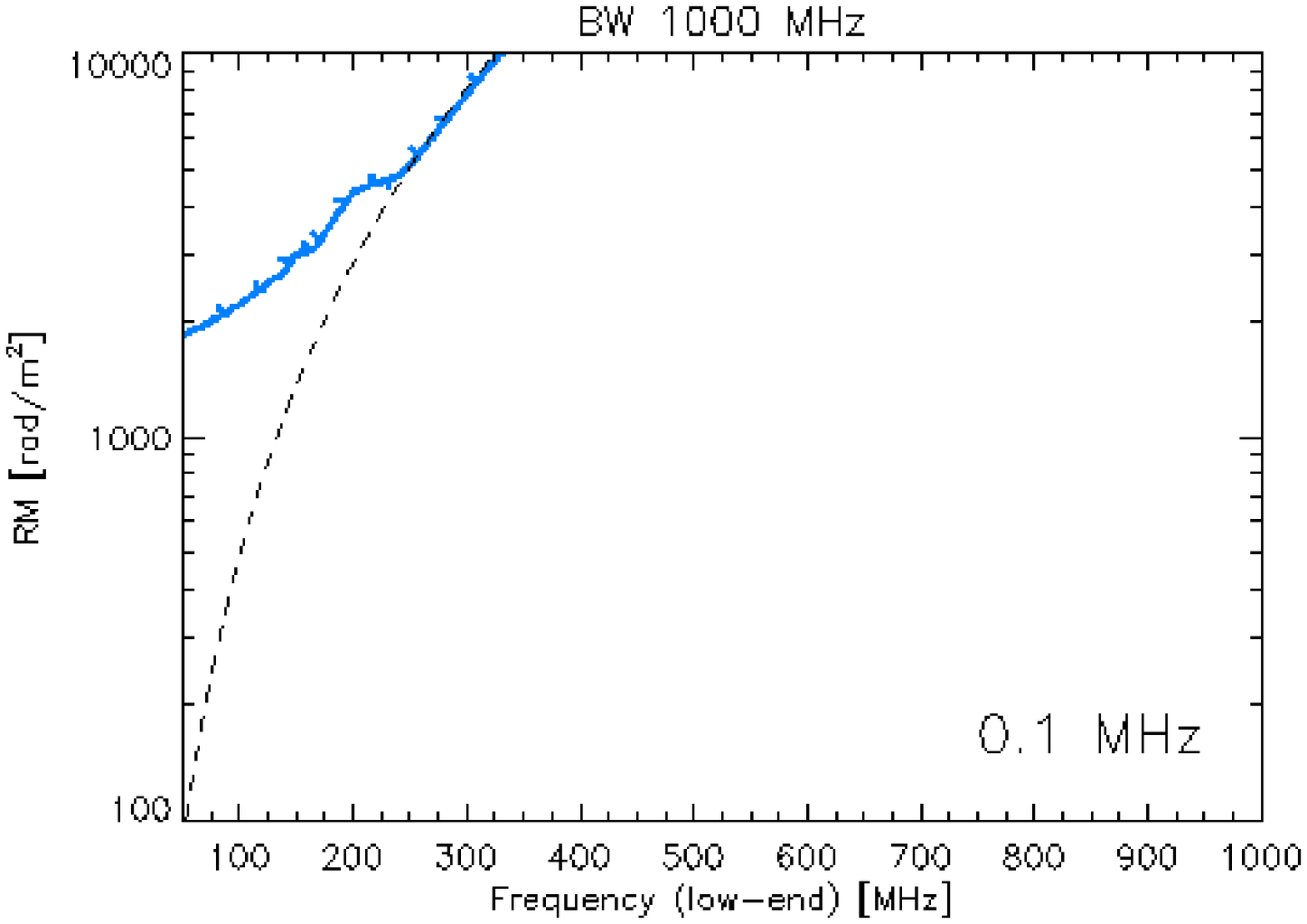}} 
    \vspace{-1ex}
%
%
    \resizebox{0.32\hsize}{!}{\includegraphics[width=\linewidth]{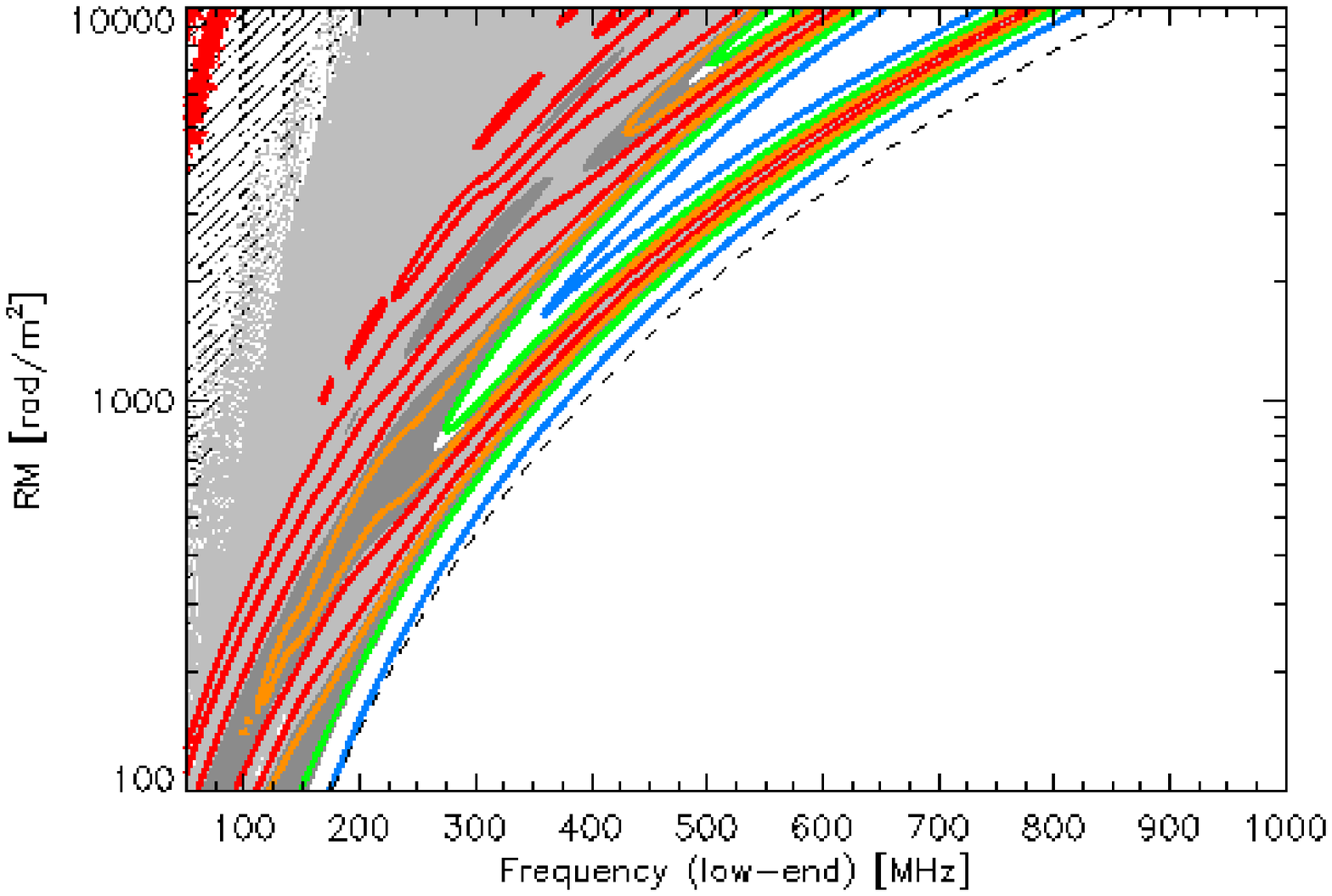}} 
    \resizebox{0.32\hsize}{!}{\includegraphics[width=\linewidth]{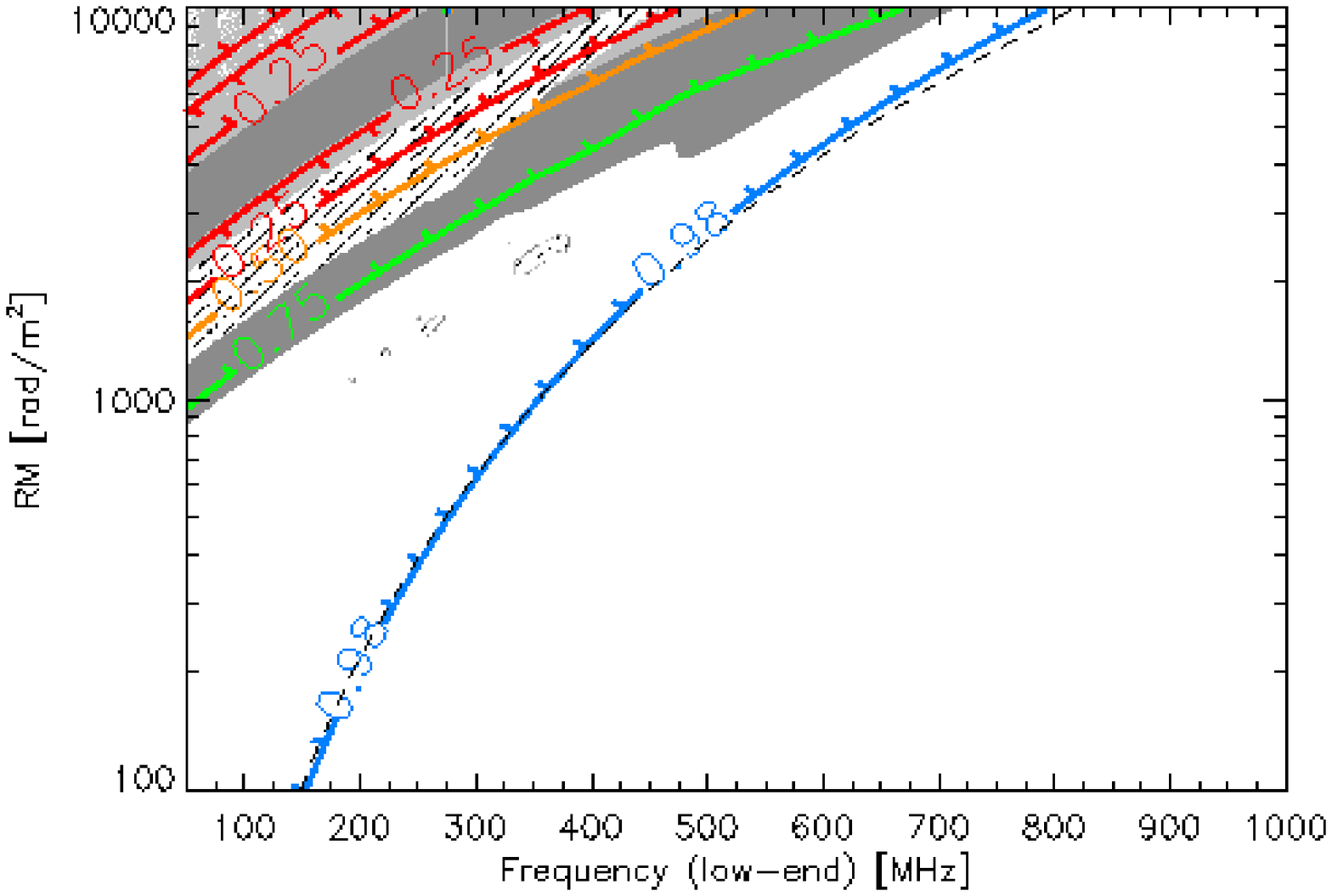}} 
    \resizebox{0.32\hsize}{!}{\includegraphics[width=\linewidth]{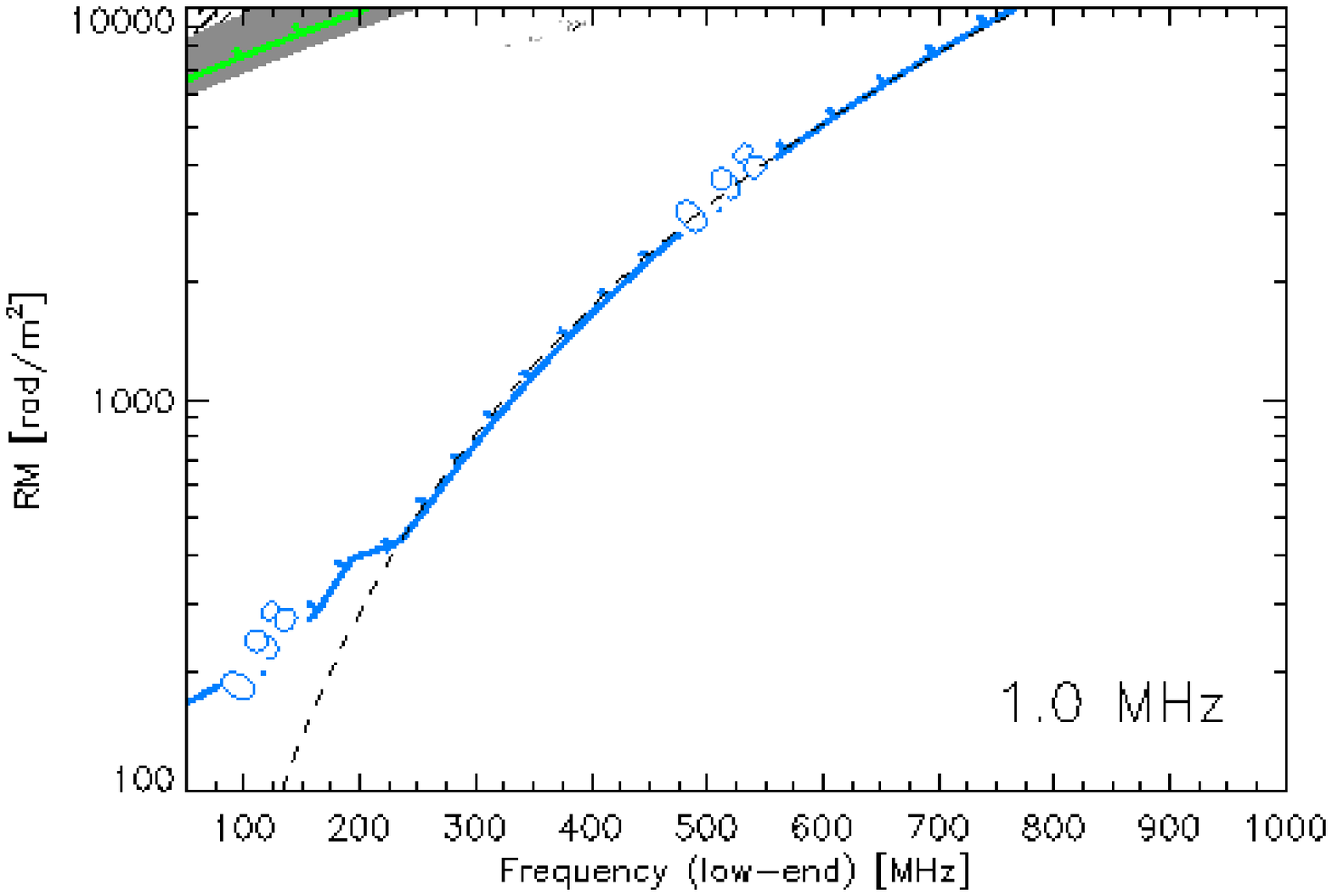}} 
\caption{
Ratio between the polarized flux density that is calculated using Equation~\ref{derot_sum} and our new formalism using the normalised version of Equation~\ref{net_derot_frequency},
each time measured at the RM at which the source emits, for a range of RMs and frequencies at the low-end of the observing band. The three columns of panels indicate bandwidths of 100 MHz, 500 MHz, and 1 GHz, respectively, while the two rows indicate frequency channels of 0.1 MHz (top row) and 1.0 MHz (bottom row). The contour lines indicate levels of 0.98, 0.75, 0.5, and 0.25, respectively. The light and dark grey backgrounds indicate when the RM spectrum that is calculated with Equation~\ref{derot_sum} is resolved, as measured by the ratio between the polarized flux density at $\mathrm{RM}'=\mathrm{RM}$ and the maximum polarized flux density in the RM spectrum. White, dark grey, and light grey backgrounds indicate ratios of $>$ 0.95, between 0.5 and 0.95, and $<$ 0.5, respectively. The hatched area indicates that the RM spectrum which is calculated using Equation~\ref{derot_sum} is heavily structured, see the text for details. The dashed line indicates the boundary given by Equation~\ref{eq:rmcon}. The short vertical line in the middle panel in the bottom row is an artefact. 
}
\label{flux_ratio.fig}
\end{figure*}

\begin{figure*}
    \resizebox{0.32\hsize}{!}{\includegraphics[width=\linewidth]{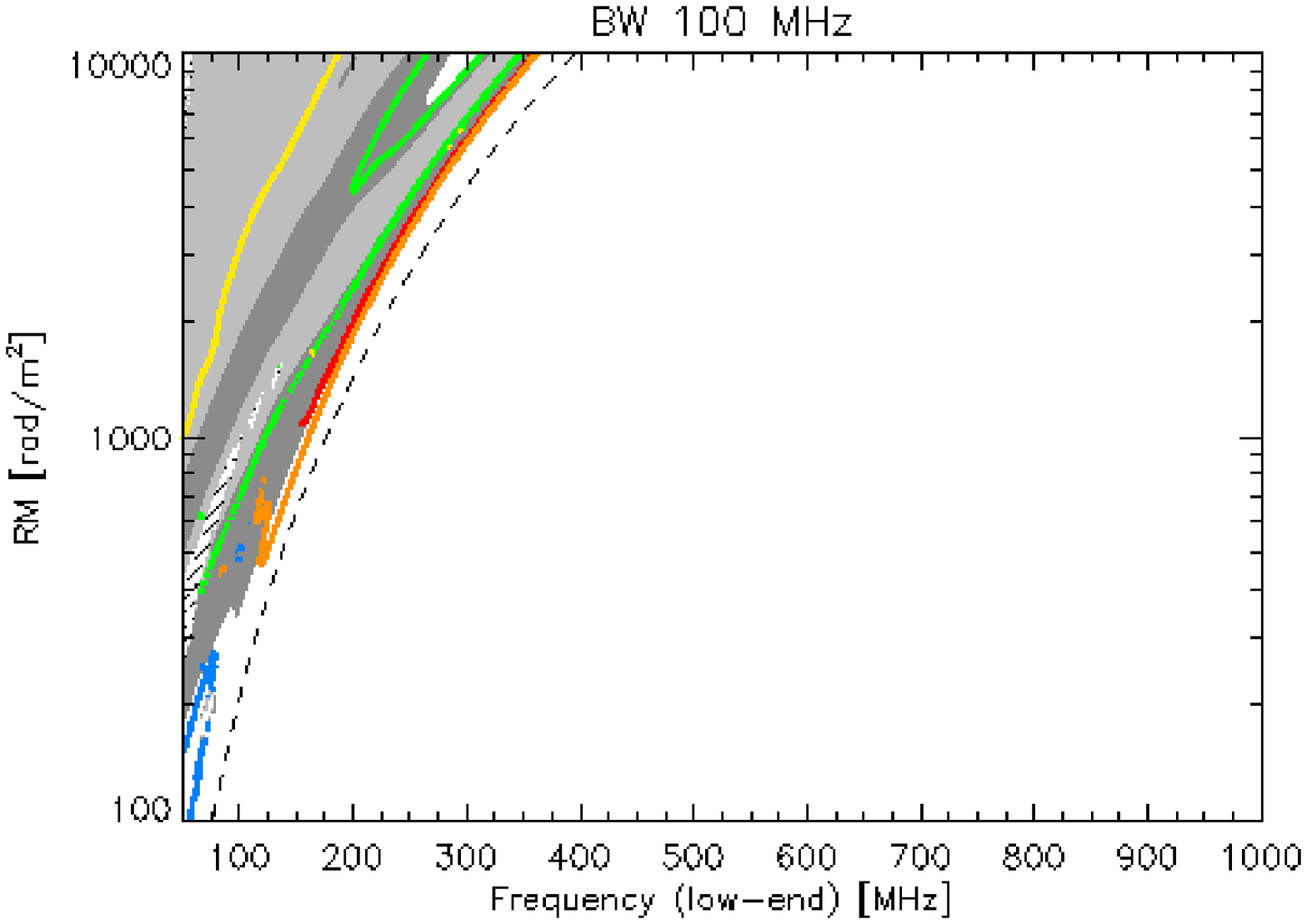}} 
    \vspace{-1ex}
    \resizebox{0.32\hsize}{!}{\includegraphics[width=\linewidth]{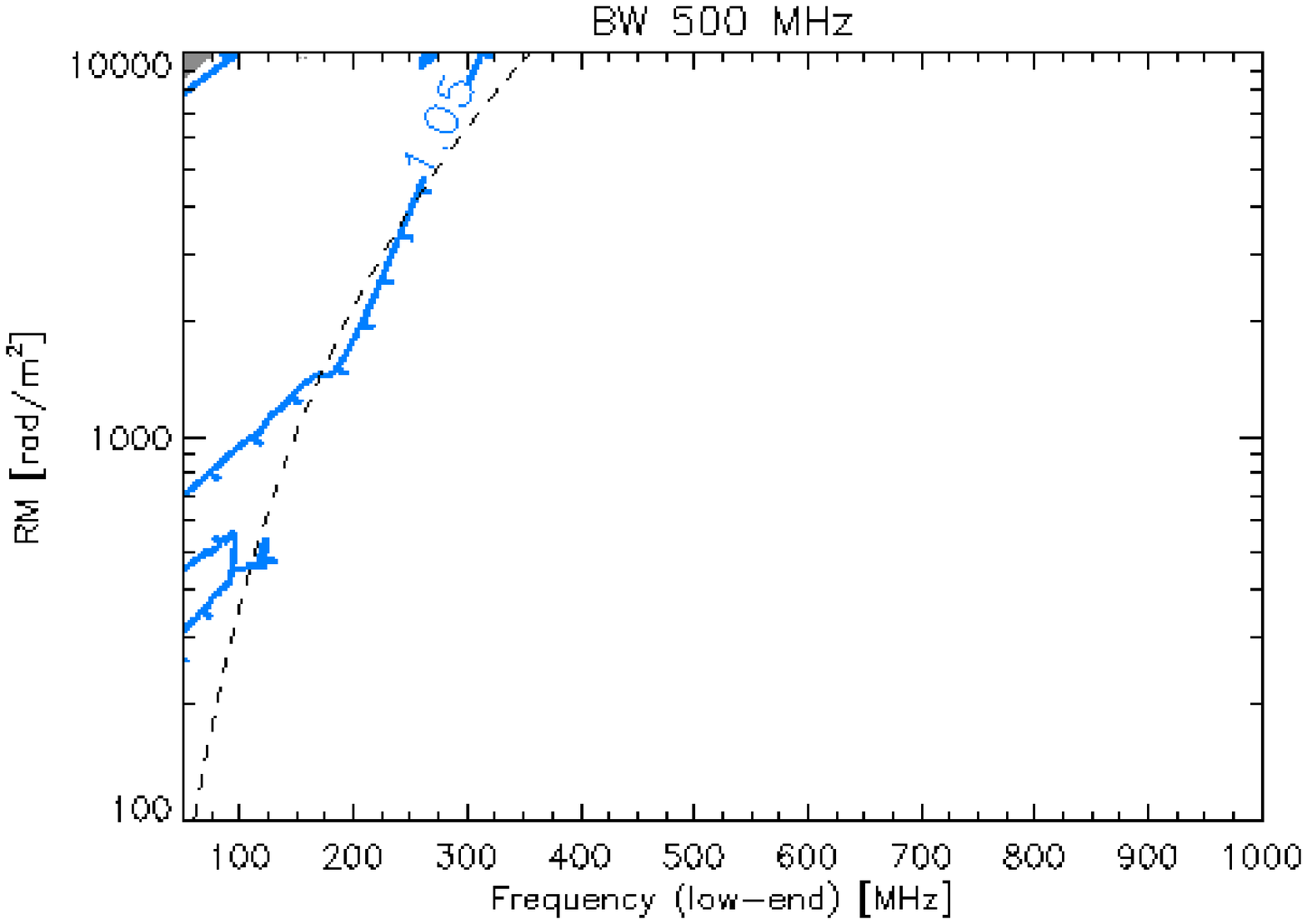}} 
    \vspace{-1ex}
    \resizebox{0.32\hsize}{!}{\includegraphics[width=\linewidth]{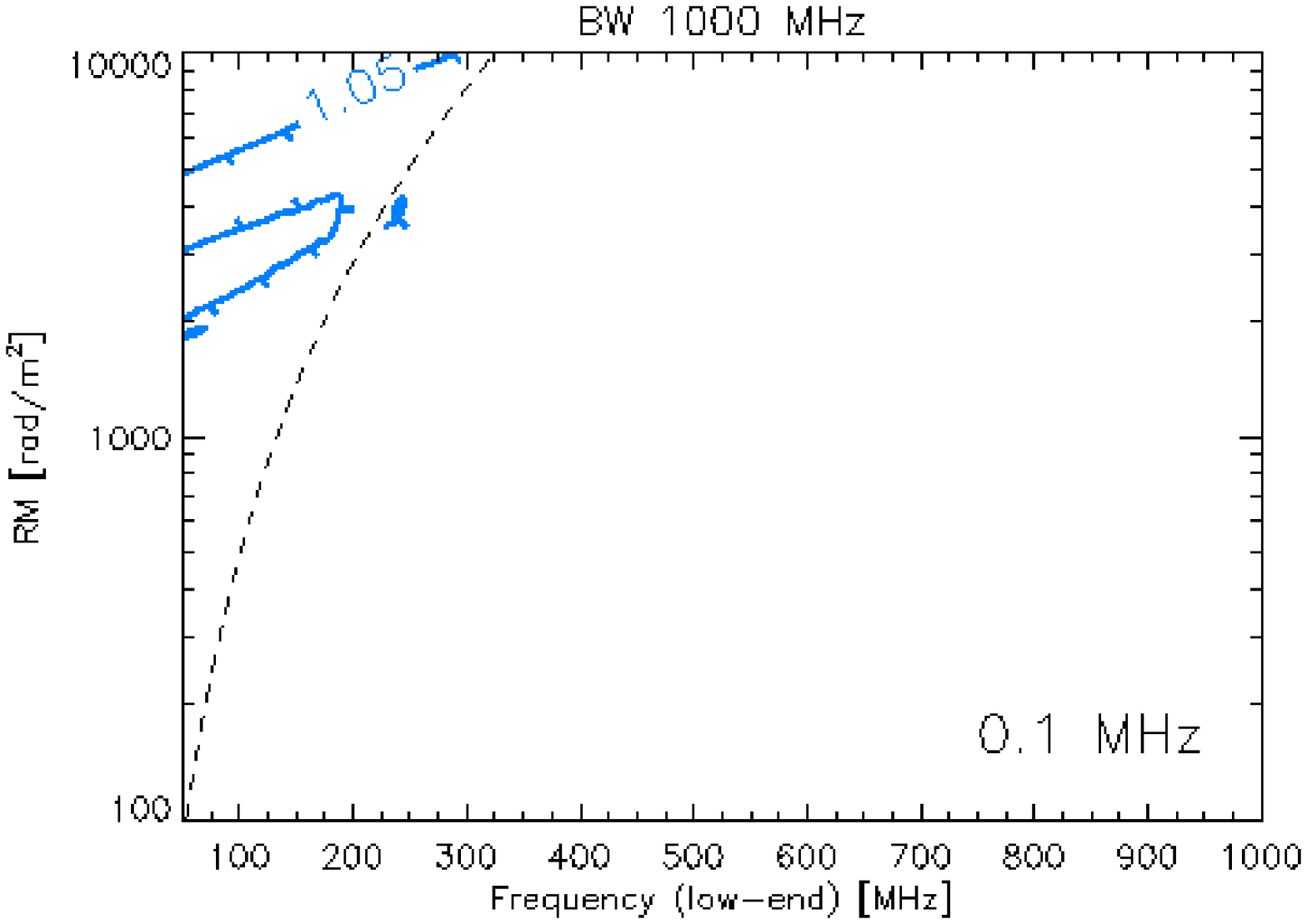}} 
    \vspace{-1ex}
%
%
    \resizebox{0.32\hsize}{!}{\includegraphics[width=\linewidth]{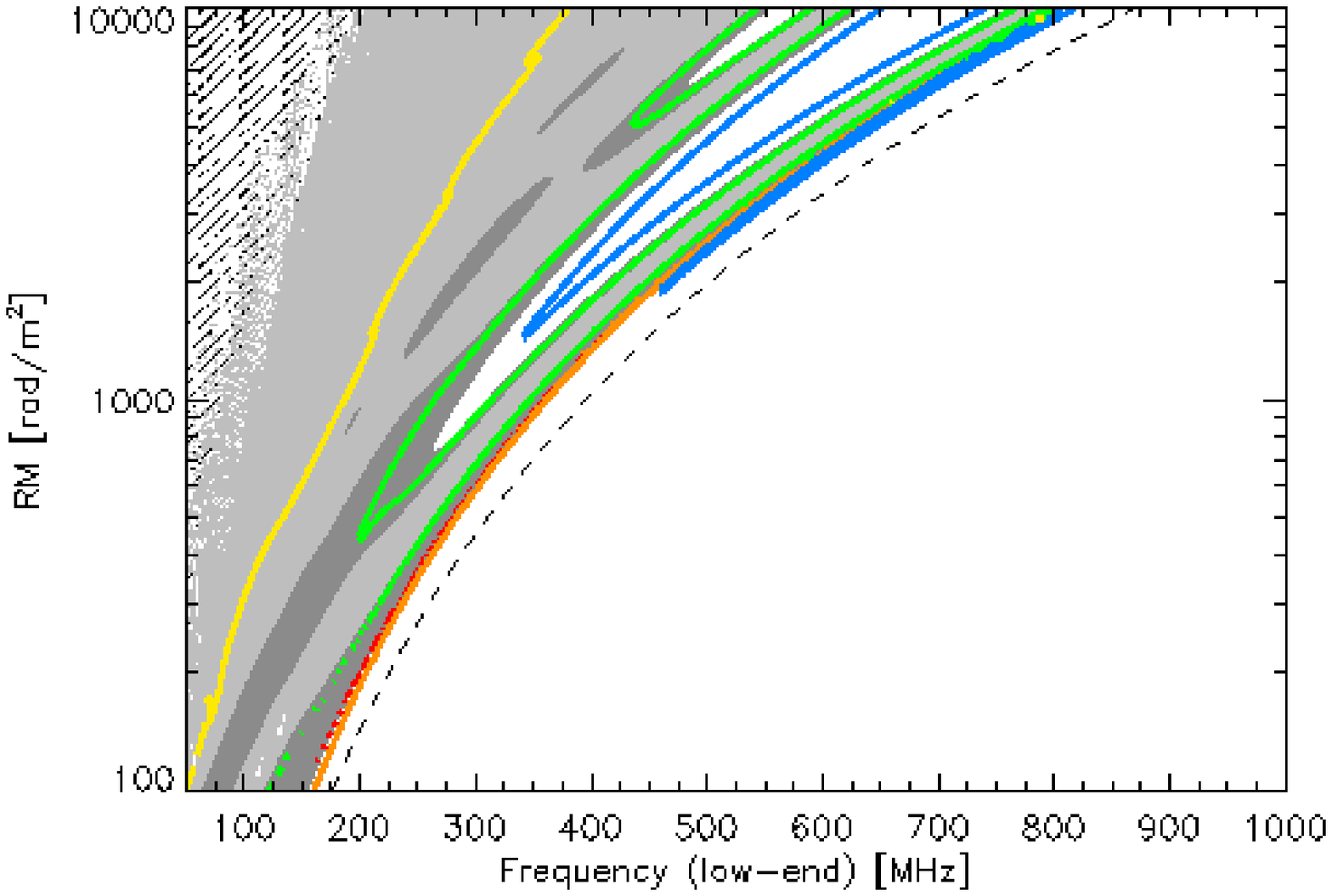}} 
    \resizebox{0.32\hsize}{!}{\includegraphics[width=\linewidth]{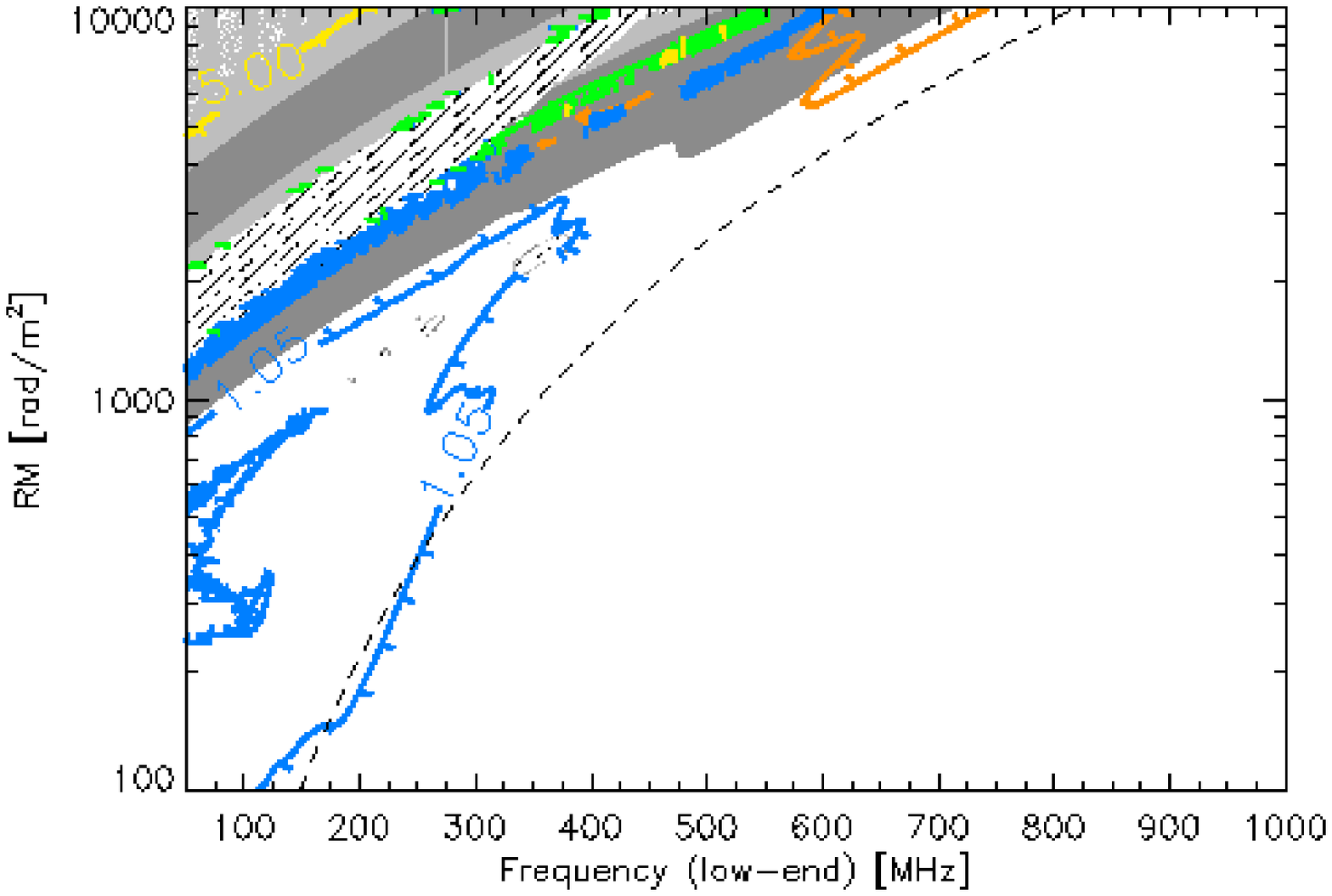}} 
    \resizebox{0.32\hsize}{!}{\includegraphics[width=\linewidth]{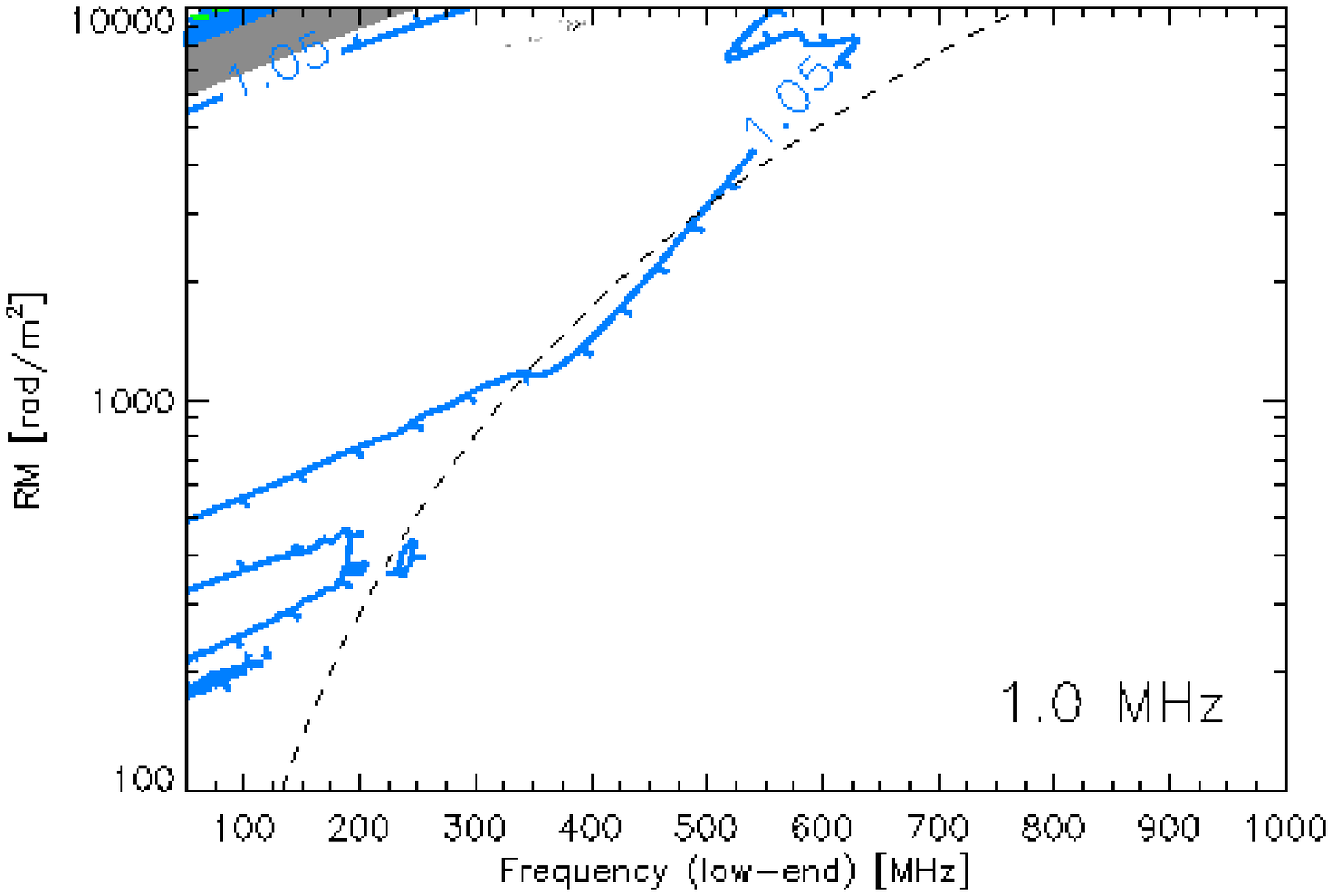}} 
\caption{The equivalent of Fig.~\ref{flux_ratio.fig} for the ratio between the FWHM of the RM spectrum that is found using Equation~\ref{derot_sum}, and the FWHM of the RM spectrum that is found when our reconstruction method is used. Contour lines indicate FWHM ratios of 0.9 (red), 0.95 (orange), 1.05 (blue), 2 (green), and 5 (yellow).
}
\label{fwhm_ratio.fig}
\end{figure*}

Fig.~\ref{RM-spec.fig} compares the RM spectra reconstructed both by Equation~\ref{derot_sum} and by our formalism. There are clear differences between the two, both in the location and depth of the RM peak about the location of the true RM value, and in certain cases it is evident that the spectrum reconstructed using Equation~\ref{derot_sum} displays a local minimum at the location of the true peak.
At these low frequencies many of the sources are severely depolarized. The Gaussian source which is observed at frequencies between 300--400 MHz shows extreme depolarization, but it is reassuring that even under such conditions both formalisms produce almost identical RM spectra.

In Figs.~\ref{flux_ratio.fig} and \ref{fwhm_ratio.fig} we elaborate on these differences by making a quantitative comparison of the two methods in the context of a model in which the background source emits at a single RM; we compare the recovered peak polarized flux densities and the FWHMs of the reconstructed spectra under the two formalisms.
Figs.~\ref{flux_ratio.fig} and \ref{fwhm_ratio.fig} capture information both on whether the RM spectrum that is found using Equation~\ref{derot_sum} is resolved (greyscale image in the background) and on the ratio between the reconstructed flux densities or FWHMs using coloured contour lines. If the RM spectrum that is found using Equation~\ref{derot_sum} becomes resolved, the FWHM of the RM spread function is calculated as the difference between the RM of the half-power point in the resolved spectrum and the RM at which the source emits. The hatched regions in Figs.~\ref{flux_ratio.fig} and \ref{fwhm_ratio.fig} indicate that the RM spectrum which is calculated using Equation~\ref{derot_sum} is heavily structured. 
This occurs either because the RM spectrum has become highly resolved and its two main peaks are ragged (panels in the bottom-left), or because the RM spectrum does not decrease monotonically to its half-power point: the FWHM then becomes ill-defined (middle panels in the bottom row).
We did not include the FWHM of these RM spectra in our analysis. 
The ratio of polarized flux densities of the two formalisms can still be calculated when the RM spectrum is resolved: therefore Figure~\ref{flux_ratio.fig} shows coloured contour lines in hatched regions.

\begin{figure}
    \resizebox{\hsize}{!}{\includegraphics[width=\linewidth]{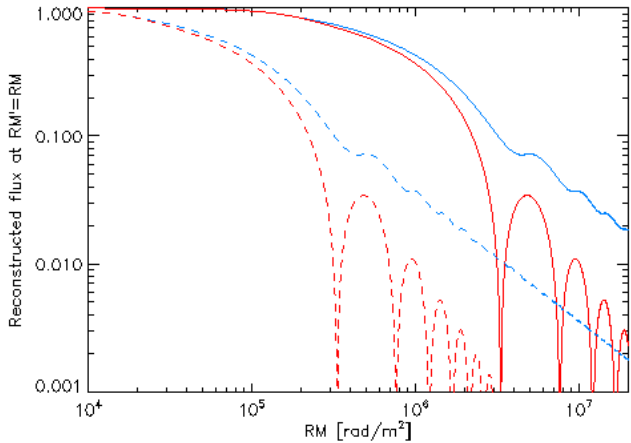}} 
\caption{Comparison of the polarized flux density that is recovered using Equation~\ref{derot_sum} and using our formalism with the normalised version of Equation~\ref{net_derot_frequency} (red and blue lines, respectively) for a source that emits one unit of flux density at one RM. The observations cover frequencies between 1 and 2 GHz with frequency channels that are either 0.1 MHz or 1 MHz wide (solid lines and dashed lines, respectively), and have a top-hat response function in frequency. 
}
\label{response.largerm.fig}
\end{figure}

The contour line in Fig.~\ref{flux_ratio.fig} where the flux density ratio is 0.98 can be used as a boundary to indicate when the two methods start to differ. We determined how this curve depends on the frequency range and channel width of the observations using a Taylor expansion of the ratio of flux densities, which is valid when the channel width $\delta\nu$ is small compared to the low-end frequency of the observing band ($\nu_\mathrm{low}$). 
This condition is met by modern radio telescopes.
We fitted for the coefficient using Fig.~\ref{flux_ratio.fig}. The boundary where the two methods start to differ can be approximated by
\begin{eqnarray}
\mathrm{RM}\approx 1.44\times 10^4\ \mathrm{rad\ m}^{-2}
\left(\frac{\nu _\mathrm{low}}{\mathrm{GHz}}\right)^{\frac{5}{2}}\left(\frac{\nu _\mathrm{high}}{\mathrm{GHz}}\right)^{\frac{1}{2}}
\left(\frac{\delta\nu}{\mathrm{MHz}}\right)^{-1}\, ,
\label{eq:rmcon}
\end{eqnarray}
\noindent
where $\nu_\mathrm{high}$ indicates the high-end frequency of the observing band. 
The RMs from Equation~\ref{eq:rmcon} can depolarize the lowest-frequency channels in the observing band by more than 90 per cent. 
Therefore Equation~\ref{derot_sum} is an excellent approximation even if depolarization at the lowest frequencies in the observing band is severe. 
Also, the condition that for Equation~\ref{derot_sum} to be valid all channels must have $\mathrm{RM}\delta\lambda^2\ll 1$ (mentioned in section~2 in B05) can be relaxed considerably.
At frequencies above 1 GHz, channel widths up to 1 MHz, and RMs up to 10,000~rad~m$^{-2}$ the differences between the two formalisms are small.

Figs.~\ref{RM-spec.fig}, \ref{flux_ratio.fig}, and \ref{fwhm_ratio.fig} show that beyond the boundary indicated by Equation~\ref{eq:rmcon} our formalism reconstructs the emitted signal more accurately than when Equation~\ref{derot_sum} is used, with a larger amplitude and in most cases a narrower FWHM of the RM spread function.
The large RMs that are involved change the orientation of the polarization vectors considerably across the frequency channels, 
and in those cases the observations can no longer be approximated by a top-hat response in wavelength squared. 
At the same time RM spectra that are calculated using our new formalism are exact because we include the response function of the simulated data;
the wings of the RM spread function can only be calculated accurately if the response function of the data is included in this calculation.

Our simulations also show that sources with large (absolute) RMs are reconstructed with a stronger signal when our new formalism is used instead of Equation~\ref{derot_sum}. 
Fig.~\ref{response.largerm.fig} illustrates this for observations between 1 and 2 GHz and frequency channels of either 1 MHz or 0.1 MHz. 
This implies that for a given sensitivity level of the observations our formalism can be used to detect sources with larger (absolute) RMs.
Fig.~\ref{response.largerm.fig} also shows that our formalism does not share the acute minima that are found using Equation~\ref{derot_sum}, which indicate RMs to which the reconstruction method is not sensitive. 
At the RM of such a minimum the ratio between the reconstructed flux densities shows a local minimum, which explains the isolated ridges in Fig.~\ref{flux_ratio.fig}.

\section{Summary}\label{sec-conclusions}
We re-formulate RM synthesis for data sets with discrete frequency channels and an arbitrary channel response function. 
The most commonly used version of the formalism by Brentjens \& De Bruyn assumes a top-hat response function in wavelength squared, which is often an approximation for the response functions of real data sets.
By including the response function of the simulated data sets the RM spectra we calculate using our formalism are exact.
We simulate mock data sets for various source geometries, using a top-hat response in frequency, and compare the properties of the RM spectra that are calculated with both formalisms. 
Equation~\ref{eq:rmcon} expresses when RM spectra that are reconstructed using the two formalisms start to differ.
We show that the formalism by Brentjens \& De Bruyn produces accurate results even if depolarization at the lowest frequencies in the observing band is severe.
The difference between the two formalisms becomes noticeable when Faraday rotation changes the orientation of the polarization vectors within channels for a large fraction of the observing band.
Under these circumstances our formalism performs better than the formalism by Brentjens \& De Bruyn.
We also show that by calculating exact RM spectra sources can be detected out to larger (absolute) RMs for a given sensitivity level of the observations.
We recommend our formalism for detecting sources with large RMs, and for understanding which RMs can be detected given an observational setup. This will be particularly important for low-frequency radio telescopes like the Low-Frequency Array (LOFAR), the Murchison Widefield Array (MWA), and the low-frequency components of the Square Kilometre Array (SKA).

\section*{Acknowledgements}
We would like to thank Bj\"orn Adebahr (Max Planck Institute for Radio Astronomy), Jean-Pierre Macquart (Curtin Institute of Radio Astronomy), and the anonymous referee for their comments that helped improve the manuscript. K.~J.~Lee gratefully acknowledges support from the ERC Advanced Grant ``LEAP'', Grant Agreement Number 227947 (PI Michael Kramer) and from the National Natural Science Foundation of China (Grant No.11373011).

\bibliography{ne6e} 

\end{document}